\documentclass[a4paper,11pt]{article}
\usepackage[latin1]{inputenc}
\usepackage{amsfonts}
\usepackage{amsmath}
\numberwithin{equation}{section}
\usepackage{amssymb}
\usepackage{amsthm}
\usepackage[T1]{fontenc}
\usepackage[dvips]{graphicx}
\usepackage{color}
\usepackage{epsfig}
\theoremstyle{plain}
\newtheorem{thm}{Theorem}
\newtheorem{lem}{Lemma}
\newtheorem{Def}{Definition}

\newtheorem{conj}{Conjecture}

\newtheorem*{rem*}{\textsc{Remark}}
\newtheorem*{def*}{Definition}
\newtheorem*{prob*}{Direct Monodromy Problem}
\newtheorem*{lem*}{\textsc{Lemma}}
\newtheorem*{cor*}{\textsc{Corollary}}
\newtheorem*{con*}{\textsc{Conjecture}}
\newcommand{\e}{\varepsilon}
\newcommand{\bb}[1]{\mathbb{ #1 }}
\newcommand{\av}[1]{\left| #1 \right|}
\newcommand{\cp}{\overline{\mathbb{C}}}
\newcommand{\beq}{\begin{equation}}
\newcommand{\eeq}{\end{equation}}

\title{Painleve I, Coverings of the Sphere and Belyi Functions}
\author{Davide Masoero \footnote{Grupo de F\'isica Matem\'atica da Universidade de Lisboa,
Av. Prof. Gama Pinto, 2
1649-003 Lisboa,  Portugal, dmasoero@gmail.com }}
\date{}
\begin{document}

\maketitle

\abstract{The theory of poles of solutions of P-I is
equivalent to the Nevanlinna problem of constructing a meromorphic function ramified over five points - counting multiplicities - 
and without critical points. We construct such meromorphic functions as limit of rational ones. In the case of the tritronquee solution they
turn out to be Belyi functions.}



\section{Introduction}
In this paper we address the following problem in the theory of Painleve equations:
given a solution $y$ of Painleve I (P-I) 
\begin{equation*}
y''(\lambda)= 6y^2(\lambda) -\lambda \, , \; \lambda \in \mathbb{C} \quad ,
\end{equation*}
and a point $w \in \cp$, characterize and compute all the inverse images of $w$, i.e.
$y^{-1}(w) \subset \bb{C}$. For definiteness we call such a problem the Inverse Painleve problem.

More precisely, we restrict to  a rather important instance of the Inverse Painleve Problem:
We study the poles of solutions of P-I. In particular we focus on the tritronquee solution, to be introduced below. 

The main idea behind the present paper is the observation \cite{piwkb} that theory of poles of solutions of P-I is
equivalent to the Nevanlinna problem of constructing a meromorphic function ramified over five points - counting multiplicities - 
and without critical points.

The first half of the paper is devoted to review this correspondence. The second half is dedicated to the effective construction
infinitely sheeted ramified coverings given the branching locus and the monodromy representation.

The correspondence between poles and the Nevanlinna problem builds on a variety of mathematical constructions which will thoroughly discuss later.
For the benefit of the reader we outline the basic structure here.
\begin{itemize}
 \item[I]Painleve-I is the equation of \textit{isomonodromic deformation} of a system of linear differential equations
with an irregular singular point \footnote{The precise form of the equation is unnecessary for the discussion. It is equation (\ref{eq:intpert}) below}.
This means that given a solution $y$ of P-I and a point $a \in \bb{C}$, then $y(a), y'(a)$ can be read from
the solution of a \textit{Riemann-Hilbert Problem}, provided $a$ is not a pole of $y$ \cite{kapaev}.

Conversely, suppose $a$ to be a pole of $y$. Then the Laurent series of $y$ looks like
\begin{equation*}
\frac{1}{(\lambda-a)^2}+\frac{a(\lambda-a)^2}{10}+ \frac{(\lambda-a)^3}{6}+b(\lambda-a)^4+O((\lambda-a)^5) \; ,
\end{equation*}
for some $b \in \bb{C}$, with all higher coefficients expressible in terms of $a,b$.

Remarkably, even though the Riemann-Hilbert problem does not have any solution, a careful study \cite{piwkb} shows that
linear system of equations converge, as $\lambda \to a$, to the cubic oscillator equation
\begin{equation}\label{eq:intcubic}
\psi''= \left(4 z^3 - 2a z - 28b \right) \psi \; .
\end{equation}

In a sentence, \textit{poles of solutions of P-I are described by the cubic oscillator (\ref{eq:intcubic}).}

\item[II]Next step is due to Nevanlinna. Given a cubic oscillator (\ref{eq:intcubic}), let $f=\psi_1/\psi_2$ the ratio of two linearly independent solutions of it.
Then $f$ is an infinitely-sheeted
branched covering of the sphere, ramified over
five points (counting multiplicities), and without any critical point.

Nevanlinna remarkably showed that the converse is true \cite{nevanlinna32}:
\textit{if a function $f$ has no critical points and it is ramified over five points, counting multiplicities, then $f$ solves, up to affine transformation of the $z$ plane,
the cubic oscillator(\ref{eq:intcubic}) for some $a,b$}.

It follows that solutions of Painleve-I are classified by the branching locus of such functions (roughly speaking five points in $\cp$ modulo the action of the M\"obius group),
while poles of a given solution are classified by equivalence classes of monodromy representations.

Remarkably, the tritronquee solution is classified by the most degenerate branching locus,
when two pairs of branching points happen to coincide and the total number of distinct branching points is three; see Theorem \ref{thm:tritrcubic} below.

\item[III]The last ingredient of the theory is the construction of infinitely-sheeted coverings of the sphere with given branching locus 
and monodromy representation. After our presentation so far, this is equivalent to solving the inverse Painleve problem for poles of solutions of P-I.

Again it was Nevanlinna \cite{nevanlinna32} who showed that they can be constructed as limit of finitely-sheeted coverings, i.e. rational functions,
with the same branching locus and increasing degree, such that their monodromy representations \textit{approximate} the given one.
\end{itemize}

The new results contained in the present paper stems from trying to make effective the Nevanlinna construction of infinitely-sheeted branched coverings as limit
finitely-sheeted one. Here is a summary.

\begin{itemize}
 \item[I] To illustrate the theory with a neat example, we compute
the sequences of rational approximants of the ratio of two solutions of $\psi''(z)=(z^2-2n-1)\psi(z), n \in \bb{N}$.
These rational functions are given two different characterizations that allow them to be computed exactly. On one hand, we give the exact location of the
critical points (see Lemma \ref{prob:fnk}), on the other hand we give a recursive formula (\ref{eq:partialschr}, \ref{eq:ncauchykharmo}). Latter characterization is still
conjectural in the general case (see Conjecture \ref{conj:fnk}), and proven in a particular case (see Theorem \ref{thm:f1k}).
\item[II] We address the Riemann problem of computing a rational function given its branching locus and the monodromy representation.
We give an algebraic treatment of that problem (see Section \ref{sec:riemann}). We show that the sought function can be described in terms of
the solution of an overdetermined system of linear equations, whose coefficients depend on the unknown location of its critical points.
After that we show that the Riemann problem can be solved by varying the coefficients of the linear systems in order to let it have a non-trivial solution.
\footnote{We expect similar ideas to have already appeared in the literature, even though we could not find any paper confirming our expectations.}
\item[III] We study the coverings of the sphere related to the poles of the tritronquee solution. We show that
the relevant monodromies are characterised by two integers $n,m$.
We identify asymptotically (i.e. for $m,n$ big enough) the location of the pole with indices $n,m$
by making a connection between the monodromy representation of cubic oscillators and
WKB analysis of the same.

Eventually, we study the rational approximation of the relevant coverings; remarkably they are all \textit{Belyi functions}, hence they are defined over
some number field. We compute explicitly some of them, in
close form, or by solving the Riemann problem numerically.
We warn the reader that our study of these rational functions is a preliminary investigation, by no means an exhaustive one.
\end{itemize}

The paper is divided into two parts: Section 2-5 are surveys of known results giving a thoroughly
exposition of the relation among poles of Painleve-I and the coverings of the sphere.
The remaining Sections contain new results. 
Section 2 introduces the Painleve-I equation and the tritronquee solution.
Section 3 is devoted to poles of solutions of P-I and the cubic oscillators.
Section 4 and 5 deals with branched coverings of the sphere related to anharmonic oscillators.
In Section 6 we illustrate the theory with a remarkable example: We study 
the rational approximants of the ratio of two solutions of $\psi''(z)=(z^2-2n-1)\psi(z), n \in \bb{N}$.
In Section 7 we address the Riemann problem of computing a rational function given its branching locus and the monodromy representation.
We give an algebraic solution of that problem.
Section 8 is devoted to constructing rational approximations of those cubic oscillators
related to poles of the tritronquee solution. 

\paragraph{Acknowledgements}
The author is since January 2012 a Scholar of the Fundacao da Ciencia e Tecnologia, scholarship number SFRH/ BPD/75908/2011.
The work started in November 2011 while the author was a Research Associate at the University of Sydney, School of Mathematics and Statistics,
partially supported by the Australian Research Council Discovery grant nos. DP0985615 and DP110102001.
We acknowledge fruitful discussions with B. Dubrovin, A. Eremenko, N. Joshi, Daniele Masoero and K. McLaughlin.
The connection between monodromy representation and WKB analysis was prompted by questions of B. Dubrovin and A. Eremenko.
We also thank A. Eremenko for kindly reading and correcting a draft of the paper.

\section{Painleve-I and the Tritronquee Solution}
Painleve I satisfies the Painleve property in a strong sense, as for any solution the only singularities in the complex plane
are poles. More precisely, any solution of P-I is a meromorphic function $y(\lambda), \lambda\in \mathbb{C}$ of order $5/2$ \cite{gromak}.

Since all solutions of P-I are meromorphic functions, then for every $y,w$ the set $y^{-1}(w)$ is countable and it does not
have any accumulation point. 

The tritronquee solution is a special solution of P-I. It was
found by Boutroux in his classical work \cite{boutroux}, where he showed that
it is the unique solution with asymptotics
\begin{equation*}
y(\lambda) \sim - \sqrt{\frac{\lambda}{6}}, \quad \mbox{if} \quad |\arg \lambda| <\frac{4 \pi}{5} \; .
\end{equation*}
The tritronquee solution may also be characterised by other analytic properties \cite{joshi} or by a Riemann-Hilbert problem \cite{kapaev}.
We will use latter characterization later on.

Recently, it has been discovered \cite{dubrovin}, and partially proven \cite{bertola10}, that the tritronquee
solution describes the leading term of the semiclassical expansion of solutions of the focusing Nonlinear Schr\"odinger Equation
close to the point of catastrophe.

In \cite{dubrovin}, the following conjecture about the location of poles of the tritronquee solution, known as \textit{Dubrovin Conjecture},
is stated.
\begin{con*}[Dubrovin]
If $\alpha \in \mathbb{C}$ is a pole of the tritronquee solution then $\av{\,\arg{\alpha}\,}\! \geq\! \frac{4 \pi}{5}$.
\end{con*}
At the time of writing the conjecture is still open \footnote{After the paper was written, a preprint with the proof of the conjecture
has been published by Costin et al. \cite{costin12}.}. However, a similar but simpler problem about
the Ablowitz-Segur solutions of Painleve-II equation has been recently solved \cite{bertola2012}.
Moreover, an analogous conjecture about absence of real poles of a solution of the Painleve-I(2) equation was proven by Claeys et al. \cite{claeys07}. 

\section{P-I and the Cubic Oscillator}
In this section we recall the relation between poles of solution of Painleve I and the cubic oscillator
\begin{equation*}
\psi''= \left(4 z^3 - 2a z - 28b \right) \psi \; .
\end{equation*}
Here we follow closely our PhD thesis \cite{myphd}.

We let 
\begin{equation} \label{eq:cubicsector}
S_k=\left\lbrace \lambda : \av{ \arg\lambda - \frac{2 \pi k}{5} } < \frac{\pi}{5} \right\rbrace
 \, , k \in \bb{Z}_{5} \; .
\end{equation}
We call $S_k$ the k-th Stokes sector.
Here, and for the rest of paper, $\bb{Z}_5$ is the group of the integers modulo five. We will often choose
as representatives of $\bb{Z}_{5}$ the numbers $-2,-1,0,1,2$.

For any Stokes sector, there is a unique (up to a multiplicative constant) solution of the cubic oscillator
that decays exponentially inside $S_k$. We call such solution the \textit{k-th subdominant solution} and let 
$\psi_k(z;a,b)$ denote it.

The asymptotic behaviour of $\psi_k$ is known explicitly in a bigger sector
of the complex plane, namely $S_{k-1}\cup\overline{S_k}\cup S_{k+1}$:
\begin{equation*}
\lim_{\substack{ z \to \infty \\ \av{\arg{z} - \frac{2 \pi k}{5}} < \frac{3 \pi}{5} -\e}}
\frac{\psi_k(z;a,b)}{z^{-\frac{3}{4}} \exp\left\lbrace-\frac{4}{5} z^{\frac{5}{2}} +
\frac{a}{2}z^{\frac{1}{2}}\right\rbrace} \to 1 , \; \forall \e >0 \, .
\end{equation*}
Here the branch of $z^{\frac{1}{2}}$ is chosen such that $\psi_k$ is exponentially small in $S_k$.

Since $\psi_{k-1}$ grows exponentially in $S_k$, then $\psi_{k-1}$ and $\psi_{k}$ are linearly independent.
Then $\left\lbrace \psi_{k-1},\psi_{k} \right\rbrace$ is a basis of solutions, whose asymptotic behaviours is known in $S_{k-1}\cup S_{k}$.

Fixed $k^* \in \bb{Z}_5$, we know the asymptotic behaviour of $\left\lbrace \psi_{k^*-1},\psi_{k^*} \right\rbrace$
only in $S_{k^*-1}\cup S_{k^*}$.
If we want to know the asymptotic behaviours of this basis in all the complex plane, it is sufficient to
know the linear transformation from basis $\left\lbrace \psi_{k-1},\psi_{k} \right\rbrace$ to basis
$\left\lbrace \psi_{k},\psi_{k+1} \right\rbrace$ for any $k \in \bb{Z}_5$.

From the asymptotic behaviours, it follows that these changes of basis
are triangular matrices: for any $k$, $\psi_{k-1}=\psi_{k+1}+\sigma_k\psi_k$
for some complex number $\sigma_k$, called \textit{Stokes multiplier}.
The quintuplet of Stokes multipliers $\sigma_k, k \in \bb{Z}_5$ is called the monodromy data of the cubic oscillator.

It is well-known (see \cite{piwkb}) that the Stokes multipliers satisfy the following system of quadratic relations
\begin{equation}\label{eq:intquadratic}
  i \sigma_{k+3}(a,b)   = 1+\sigma_k(a,b)\sigma_{k+1}(a,b)\, , \; \forall k \in \bb{Z}_5 \,,\; \forall a,b \in \bb{C} \; .
\end{equation}

Hence, it turns out that the monodromy data of any cubic oscillator is a point of a two-dimensional smooth algebraic subvariety of
$\bb{C}^5$ cut by \ref{eq:intquadratic}, called \textit{space of monodromy data}, which we denote by $M_5$.
\paragraph{Poles of Solutions of P-I}
It is well-known that P-I is the equation of \textit{isomonodromic deformation} of a system
of linear differential equations with an irregular singular point. The choice
of the linear equation is not unique, see for example \cite{takei}, \cite{kapaev}.

Here we follow \cite{takei} and choose the following auxiliary equation
\begin{eqnarray}\label{eq:intpert}
 \frac{d^2\psi(z)}{dz^2} \!\! &=& \!\! Q(z;y,y',\lambda) \psi(z)\; , z,y,y',\lambda \in \bb{C} \\ \nonumber
 Q(z;y,y',\lambda) \!\! &=& \!\! 4 z^3 - 2 \lambda z + 2 \lambda y - 4 y^3 + y'^2+  \frac{y'}{z - y} 
+\frac{3}{4(z -y)^2} \, .
\end{eqnarray}
We call such equation the \textit{perturbed cubic oscillator}.

It turns out that one can define subdominant solutions $\psi_k$, and Stokes multipliers $\sigma_k, k\in \bb{Z}_5$
also for the perturbed cubic oscillator. Moreover, also the Stokes multipliers of the perturbed oscillator
satisfy the system of quadratic relations (\ref{eq:intquadratic}); hence, the quintuplet of Stokes multipliers of any perturbed
cubic oscillator is a point of the space of monodromy data $M_5$.

Since P-I is the equation of isomonodromy deformation of the perturbed cubic oscillator \footnote{
Let the parameters $y=y(\lambda), y'=\frac{dy(\lambda)}{d\lambda}$ of the potential $Q(z;y,y',\lambda)$
be functions of $\lambda$; then $y(\lambda)$
solves P-I if and only if the Stokes multipliers of the perturbed oscillator do not depend on $\lambda$}
we can define a map $\cal{M}$ from the set of solutions of P-I to
the space of monodromy data; fixed a solution $y^*$,  ${\cal{M}}(y^*)$ is
the monodromy data of the perturbed cubic oscillator with potential $Q(z;y^*(\lambda),y^{*}~'(\lambda),\lambda)$,
for any regular value $\lambda$ of $y^*$ . The map $\cal{M}$ is a special
case of Riemann-Hilbert correspondence \cite{kapaev}.

In this paper we are mainly interested in studying poles of solutions of P-I. To this aim
we cannot use directly the perturbed
oscillator in this study because the potential
$Q(z;y,y',\lambda)$ is not defined at poles, i.e. when $y=y'=\infty$.

However, this problem can be overcome by studying the auxiliary equation in the proximity of a pole.
In fact the auxiliary problem has a well defined limit.

\begin{lem}\label{lem:limittopole}
Let $a$ be a pole of a fixed solution $y^*(\lambda)$ of P-I and let $\psi_k(z;\lambda)$
denote the k-th subdominant solution of the perturbed cubic oscillator (\ref{eq:intpert}) with potential $Q(z;y^*(\lambda),y'^*(\lambda),\lambda)$.
In the limit $z \to a$,
$\psi_k(z;\lambda)$ converges (uniformly on compacts) to the k-th subdominant solution $\psi_k(z;a,b)$
of the cubic oscillator
\begin{equation*}
\psi''= \left(4 z^3 - 2a z - 28b \right) \psi \; .
\end{equation*}
Here the parameter $b$ is the coefficient of the $(z-a)^4$ term in the Laurent expansion
of $y^*$: $y^*=\frac{1}{(\lambda-a)^2}+\frac{a(\lambda-a)^2}{10}+ \frac{(\lambda-a)^3}{6}+b(\lambda-a)^4+O((\lambda-a)^5)$.
\begin{proof}
See \cite{myphd} Chapter 4 or \cite{piwkb}.
\end{proof}
We remark that a similar Lemma for poles of solutions of Painleve-II was proven in \cite{its}.

\end{lem}
As corollaries of the previous Lemma we were able to prove the following Theorems,
which define precisely the relation between P-I and the cubic oscillator.
\footnote{Even though the statement
of Theorems \ref{thm:polescubic}, \ref{thm:tritrcubic} already appeared
in \cite{chudnovsky94} by D. Chudnovsky and G. Chudnovsky, in \cite{piwkb} we gave a
(perhaps the first) rigorous proof \cite{piwkb}. Theorem \ref{thm:rhcorr}
can be proven also by other means \cite{elliptic}.}.

\begin{thm}\label{thm:polescubic}
Fix a solution $y^*$ and call $\sigma_k^*, k \in \bb{Z}_5$ its Stokes multipliers:
${\cal{M}}(y^*)=\left\lbrace\sigma_{-2}^*,\dots,\sigma_2^*\right\rbrace$.

The point $a \in \bb{C}$ is a pole of $y^*$ if and only if there exists $b \in \bb{C}$
such that $\sigma_k^*, k \in \bb{Z}_5$ are Stokes multipliers of the cubic oscillator
\begin{equation*}
\psi''= \left(4 z^3 - 2a z - 28b \right) \psi \; .
\end{equation*}
The parameter $b$ turns out to be the coefficient of the $(\lambda-a)^4$ term in the Laurent expansion
of $y^*$. 
\end{thm}
\begin{thm}\label{thm:tritrcubic}
Poles of integrale tritronqu\'ee are in bijection with cubic oscillators
 such that $\sigma_2=\sigma_{-2}=0$. In physical terminology, these cubic oscillators are said to
 satisfy two "quantization conditions".
\end{thm}
\begin{thm}\label{thm:rhcorr}
 The Riemann-Hilbert correspondence $\cal{M}$ is bijective.
 In other words, $M_5$ is the moduli space of solutions of P-I.
\end{thm}
The proofs of above Theorems can be found in \cite{piwkb} and in \cite{myphd}.
\section{Branched Coverings of the Sphere}
Here we introduce some notions about branched coverings of the sphere in order to be able to classify
poles of solutions of Painleve-I. Since we deal also
with coverings from non compact surfaces, the material is not that standard. Our main sources are \cite{eremenko2004}, \cite{nevanlinna}.

Given two (not necessarily compact) Riemann surfaces $M$ and $N$, we say that $\varphi: M \to N$ is a branched covering if
$\varphi$ is a non-constant holomoprhic map and if there is a finite set $B \subset N$ such
that $\varphi: M \setminus \varphi^{-1}(B) \to N \setminus B$
is a topological covering. The smallest of all sets $B$ is called the \textit{branching locus}.
The elements of the branching locus are called \textit{branch points}.

Two maps $\varphi,\chi$ from $M$ to $N$ are said equivalent if there is an automorphism $\alpha$ of $M$
such that $\varphi=\chi \circ \alpha$.

In this paper the target surface will always be $\cp$, while the domain will be either $\bb{C}$ or $\cp$.
We recall that the automorphism of $\bb{C}$ are the affine maps, $z \to az+b, a\neq 0$, while
the automorphisms of the sphere are the fractional linear (or M\"obius) transformation $z \to \frac{az+b}{cz+d}, ab-cd \neq 0$. 

Let $B_f$ be the branching locus of $f$.
Chosen a point $w$ of the punctured sphere $\cp \setminus B$ and a numbering of $f^{-1}(w)$ (a bijection from $f^{-1}(w) \to \bb{Z}_n$, $n$ possibly $\infty$),
we obtain a representation $\rho$ of the fundamental group $\pi(\cp \setminus B_f, w)$ in the group permutations of $n$ elements, $S_{n}$.
This is called the monodromy representation and it is defined up to numbering of the set $f^{-1}(w)$,
i.e. up to conjugations in $S_n$.
\subsection{Rational functions}
Holomorpic functions from the sphere to itself are the rational functions and every rational function defines a
branched covering of the sphere. 

Given a rational function $f$, $z^*$ is a critical point of $f$ if the differential of $f$ evaluated at $z^*$ is zero.
In a neighborhood of a critical point $z^{*}$, choosing appropriate local coordinates,
$f$ has the normal form $\tilde{f}(z)=(z-z^{*})^{k}$, $k\geq2$. The multiplicity of the critical point ${z^{*}}$ is  $k-1$.
A point $b^{*}$ is a critical value if it is the value of $f$ at one of its critical points.
The branching locus of $f$ is the totality of its critical values.

Given $f$ we can read the number of critical points and their multiplicities from the monodromy representation.
Let $\gamma_b$ be a small circle around the critical value $b$, then its image $\rho(\gamma_b)$ in $S_n$
is the product of some distinct cycles. Every non trivial cycle correspond to a critical point and
the multiplicity of that critical point is equal to the length of the corresponding cycle minus one. 

The number of critical points (counting multiplicities) are related to the degree of the function
by a simple relation called Riemann-Hurwitz formula:
Let $z_{1},\dots,z_{m}$ be the critical points of $f$ and let $v_{1},\dots,v_{k}$ the respective multiplicities.
Then
\beq \label{eq:riem-hurw}
\deg{f}=1+\frac{\sum_{k=1}^{m}v_{k}}{2}
\eeq
Notice that the Riemann-Hurwitz formula is a non-trivial relation in the cycle structure of the monodromy representation of $f$.

Equivalence classes of rational functions are in bijection with all monodromy representations satisfying
the Riemann-Hurwitz formula. This is the celebrated \textit{Riemann Existence Theorem}, which we take from \cite{catanese95}.
\begin{thm}[Riemann's Existence Theorem]
Fix a finite set $B \subset \cp $ and a morphism $\rho: \pi(\cp\setminus B,x) \to S_n$, for some finite $n$
such that
\begin{itemize}
 \item[(i)] the image of $\pi(\cp \setminus B,x)$ is a transitive subgroup of $S_n$
\item[(ii)] for any $b \in B$, if $\gamma_b \in \pi(\cp \setminus B)$ is (conjugate to) a small circle around $b$ then
$\rho(\gamma_b)$ is not the identity. 
\item[(iii)]the monodromy representation satisfy the Riemann-Hurwitz formula (\ref{eq:riem-hurw}).
\end{itemize}
There exists a rational map $f$, unique up to equivalence, whose branching locus is $B$ and whose monodromy representation is $\rho$.

Hence equivalence classes of rational functions whose branching locus is $B$ are in bijection with conjugacy classes of monodromy representations
satisfying (i,ii,iii).
\end{thm}

\subsection{Meromorphic functions}
Holomorphic functions from the complex plane to the sphere are called meromorphic functions. Notice that
not every meromorphic function is a branched covering of the sphere according to our definition.

In what follows by a meromorphic function we always mean a transcendental one. Following closely \cite{eremenko95}
we introduce the concept of \textit{transcendental singularity}.

For any $a \in \cp$ and disc $D(r,a)$ of radius $r>0$ centered in $a$, let $U(r)$ be a component
of $f^{-1}(D(r,a))$ chosen in such a way that $U(r_1) \subset U(r_2)$ for any $r_1<r_2$.
Then either (a) or (b) holds:
\begin{itemize}
 \item[(a)]$\cap_{r>0}U(r)=z, z\in \bb{C}$ and $f(z)=a$. $z$ is either an ordinary point or a critical point of $f$.
\item[(b)]$\cap_{r>0}U(r)=\emptyset$. Then the map $r \to U(r)$ is a \textit{transcendental singularity} of $f^{-1}$.
We say that the singularity lies over $a$. A simple argument \cite{eremenko95} shows that $a$ is an \textit{asymptotic value} of $f$, i.e. $a$ is the limit
of the function $f$ evaluated on a continuous curve tending to $\infty$.
According to the \textit{Iversen classification}, the transcendental singularity is either 
\item[(bi)]\textit{direct}, if there exists $r$ such that $f(z) \neq a$ for any $z \in U(r)$. In this case $f:U(r) \to D(r,a)\setminus a$ is the universal cover
of $D(r,a)\setminus a$. Or
\item[(bii)]\textit{indirect}, otherwise.
\end{itemize}

Due to Hurwitz Theorem \cite{eremenko2004} we have that $f:f^{-1}(D) \to D$ is a covering map if and only if $D$ contains no
critical value and no asymptotic values. Then,
according to our definition, $f$ is a branched covering if and only if the union of the critical values and of the asymptotic values is a finite set.

In this class of functions, the most important Theorem.

\begin{thm}[Nevanlinna, Elfving]\label{thm:prenevanlinna}
Let $f$ be a function with $p <\infty$ transcendental singularities and $m <\infty$ critical points, lying over $q \geq 2$ points.
Then all transcendental singularities are direct and
\begin{equation}\label{eq:thmschwarzian}
 \left\lbrace f(z) , z \right\rbrace \equiv \frac{f'''(z)}{f'(z)} -
\frac{3}{2}\left(\frac{f''(z)}{f'(z)}\right)^2 = r(z) \; ,
\end{equation}
where $r$ is a rational function of degree less or equal than $p-2+2m$.

In particular, suppose the function $f$ has no critical points, i.e. $m=0$. Then 
$r(z)$ is a polynomial of degree $p-2$.
\begin{proof}
For the case $m=0$ see \cite{nevanlinna32}. For the general case, see \cite{elfving}.
\end{proof}
\end{thm}
The operator $ \left\lbrace f(z) , z \right\rbrace$ is called the \textit{Schwarzian derivative}.
Formula (\ref{eq:thmschwarzian}) shed some light on the connection between branched coverings of the sphere and linear differential equations of the second order.
Indeed, it is well-known that a function $f$ satisfy $\left\lbrace f(z) , z \right\rbrace=k(z)$ if and only if
$f=y_1/y_2$ where $y_1,y_2$ are two linearly independent solutions of $y''(z)=-2 k(z)y(z)$.

Conversely if a function satisfy (\ref{eq:thmschwarzian}) with $k(z)$ being a rational function then $f$ has a finite number of critical points and
transcendental singularities (all being direct). We show this below in the case, the only relevant for us, where $k(z)$ is a polynomial.
 
\subsection{Anharmonic oscillators}\label{sec:nevanlinna}
By anharmonic oscillator we mean a Schr\"odinger equation with a polynomial potential $V_m$ of degree $m \geq 1$
\begin{equation}\label{eq:manharmonic}
 y''(z)=V_m(z) y(z)   \; .
\end{equation}
Equivalently we mean a meromorphic function whose Schwarzian derivative is a polynomial.
In fact a function satisfy
\begin{equation}\label{eq:schwarzian}
 \left\lbrace f(z) , z \right\rbrace \equiv \frac{f'''(\lambda)}{f'(\lambda)} -
\frac{3}{2}\left(\frac{f''(\lambda)}{f'(\lambda)}\right)^2 =-2 V_m(z) \; ,
\end{equation}
if and only if it is the ratio of two independent solutions of (\ref{eq:manharmonic}).
Particular cases of anharmonic oscillators are the Airy equation $m=1$, the harmonic oscillator $m=2$ and the cubic oscillator $m=3$.

Here, and for the rest of the paper if not otherwise stated, we always suppose that the polynomial $V_m$ is monic and normalized as follows
\begin{equation*}
V_m(z)=z^m+ a_2 z^{m-2}+ \dots + a_m \, , \; m\geq 1 \; .
\end{equation*}

Following \cite{eremenko}, we outline the coverings structure of the general solution of the anharmonic oscillator.
With this normalization, the Stokes sectors are defined as
\begin{equation*}
S_k=\left\lbrace \lambda : \av{ \arg\lambda - \frac{2 \pi k}{m+2} } < \frac{\pi}{5} \right\rbrace
 \, , k \in \bb{Z}_{m+2} \; .
\end{equation*}
Let $f$ be a solution of (\ref{eq:schwarzian}). Then
$f$ satisfies several remarkable properties
\begin{itemize}
\item $f$ has no critical points.
 \item Along any ray $r=\rho e^{i\varphi}, \rho \in \bb{R}^+$ contained in $S_j$ 
\begin{equation}\label{eq:wj}
 \lim_{\rho \to \infty} f(r) = w_j \; .
\end{equation}
Moreover, $\lbrace w_0, \dots, w_{m+1}\rbrace$ are all the asymptotic values of $f$. Hence $f$ has $m+2$ transcendental singularities (all direct) and no critical points.
\item $w_j \neq w_{j+1}$ for any $j \in \bb{Z}_{m+2}$
\end{itemize}

Due to Theorem \ref{thm:prenevanlinna} and the previous discussion, we have the following Theorem \cite{nevanlinna32}.
\begin{thm}[Nevanlinna]\label{nevanlinnaeremenko}
A meromorphic function $g$ has $m+2$ transcendental singularities and no critical points if and only
$g=f(az+b)$, $\left\lbrace f(z) , z \right\rbrace = V_m(z)$, for some $a,b \in \bb{C}, a \neq 0$ and some $V_m$ of degree $m$.
\end{thm}
Thanks to Nevanlinna theory, we are now able to characterize poles of the integrale tritronquee in terms of branched coverings of the sphere. Indeed
from Theorem \ref{thm:tritrcubic} we know that poles of integral tritronquee
are in bijection with cubic oscillators such that $\sigma_2=\sigma_{-2}=0$.
The relation $ \sigma_2=\sigma_{-2}=0$ is readily translated into a relation among the asymptotic values
\footnote{The general transformation between Stokes multipliers and asymptotic values was given in \cite{dtba}. We will not need it here.}: in fact it is equivalent to require
that two pairs of asymptotic values coincide, namely $w_1=w_{-2},w_{-1}=w_2$; here the $w_j$ are defined by (\ref{eq:wj}).
Hence, poles of the integrale tritronquee are classified by coverings of the sphere
with five transcendental singularities lying over three points and no critical points. More precisely we have the following
\begin{thm}\label{thm:polesvalues}
Poles of integrale tritronquee are in one-to-one correspondence with (equivalence classes) of branched coverings of the sphere $f:\bb{C} \to \cp$,
with five transcendental singularities lying over $0,1,\infty$ and no critical points.

If we normalize the covering map $f$ in such a way that $\lbrace f,z \rbrace =4 z^3 - 2 a z -28 b $ and
$w_0=0,w_1=w_{-2}=1,w_2=w_{-1}=\infty$. Then $a$ is a pole of the integrale tritronquee. 
\begin{proof}
Theorem \ref{thm:tritrcubic} and Theorem \ref{nevanlinnaeremenko}
\end{proof}
\end{thm}
In Section \ref{sec:polesriemann} below, we will classify the branched coverings related to the poles of the tritronquee by exhibiting
explicitly their monodromy representations. To this aim, we introduce some elements of the combinatorial theory of coverings.
\section{Combinatorics of Coverings}
In this section we introduce some elements of the combinatorial theory of branched coverings of the sphere.
Here we follow mainly \cite{nevanlinna}, \cite{elfving}. As an application of this theory, we will show (following \cite{nevanlinna32}) how
infinitely-sheeted coverings can be constructed as limit of finitely-sheeted ones.

In this section, if not otherwise stated, we suppose that the branching locus $b_1,\dots,b_n $ is given and ordered and that we have fixed an oriented Jordan curve $\gamma$
passing through all the branching points, respecting the ordering. This curve divides the sphere into two polygons,
the inner and outer polygon with sides the arcs $(b_1,b_2),\dots (b_n,b_1)$.

Let $f$ have branching points $b_1,\dots,b_n$, we call Critical Graph of $f$ the inverse image of the curve $\gamma$ and of the branch points $b_1,\dots b_n$ \cite{nevanlinna}.

The Critical Graph allows us to read the critical structure of $f$. In fact,
a critical point of $f$ is a vertex of the critical graph whose valence is bigger than 2.
The multiplicity of a critical point is half the number of those faces minus one. The valence of the vertex is also the number of the distinct
polygons meeting at the vertex. Similarly, a direct transcendental singularities is an asymptotic direction of the complex plane where infinitely many
polygons meet: If we map the plane to the unit disc, they become vertex lying on the unit circle whose valence is infinite.

In those cases where the critical points lie on the circle of radius 1, then we always choose $\gamma$ to be the circle of radius one:
The Critical Graph is simply the union of the level curves $|f(z)|^2=1$.

The dual of such a picture is called \textit{line complex} \cite{nevanlinna}. Choose a point in the inner polygon $P_i$ and a point $P_o$ in the outer polygon, and
for any side $(b_k,b_{k+1})$ a line $l_k$ connecting $P_i$ and $P_o$ crossing just the side $(b_k,_k{k+1})$ and only once (such a line
is unique up to homotopy in $\cp  \setminus \lbrace b_1,\dots,b_n \rbrace$).
The line complex is the inverse image of the union of the lines $l_k$. The vertices are bipartite being either inner (inverse images of $P_i$)
or outer (inverse images of $P_o$).

From the line complex we can easily compute the monodromy representation. For any $k$, we define a map from the inner vertices to the outer
vertices (and its inverse): $\mu_k(v)$ is the vertex adjacent to $v$ w.r.t. the edge $l_k$.
The composition  $\Sigma_k(v)=\mu_{k}\circ \mu_{k-1}(v)$ of two maps is a permutation of the inner (or of the outer) vertices. It
defines the monodromy representation of a circle around the branch point $b_k$ and base point $P_i$ (or $P_o$).

Notice that the line complex of $f$ (up to orientation-preserving homeomorphism of the domain) is far from being unique if $n>2$,
as it depends on the cyclic ordering of the branch points that we have chosen.
In fact the braid group acts transitively on the possible line complexes of a function; see \cite{lando2004}
for the case of rational functions, and \cite{eremenko} for the case of meromorphic functions.

\paragraph{An Example}
We consider the rational function $f=\frac{1+3z-3z^2+3z^3}{3-3z+3z^3+z^3}$.
It has three critical points $1,i,-i$ with multiplicity $2,1,1$ and critical value $1,i,-i$.
Below we show on the left its line complex and on the right its Critical Graph. Since the critical values lie on the circle
of radius $1$, we choose the circle to be the Jordan curve passing through the critical values.
In purple is $|f(z)|<1$, in grey $|f(z)|>1$.

\begin{minipage}[c]{.40\textwidth}
\begin{center}
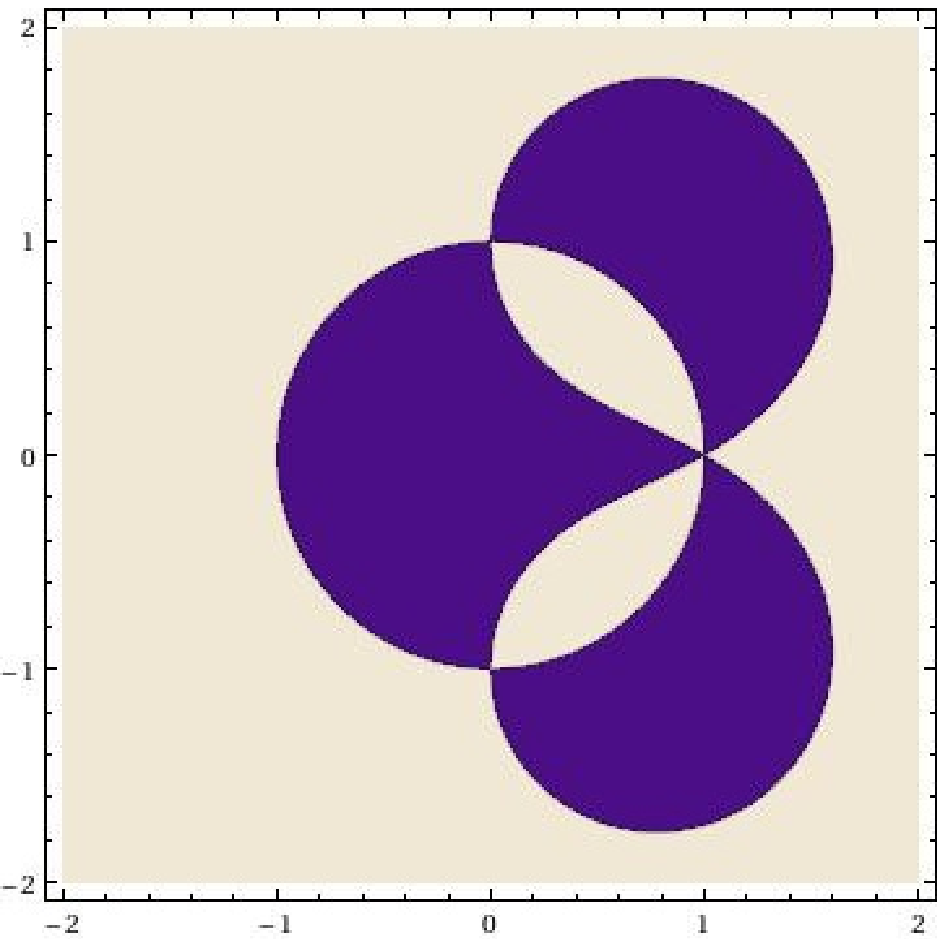
\end{center}
\end{minipage}
\begin{minipage}[c]{.40\textwidth}
\begin{center}
\includegraphics[width=4cm]{simplest.eps}
\end{center}
\end{minipage}

To the aim of classifying coverings of the sphere, we formalise the notion of a line complex (for more details, see \cite{goldberg70})
and we show for any line complex on $\bb{C}$ (resp. $\cp$)
there is a branched coverings from  $\bb{C}$ (resp. $\cp$) to $\cp$ with that diagram.
In other words, we can restate Riemann Existence Theorem and Theorem
\ref{nevanlinnaeremenko} in terms of line complexes.

An abstract line complex of index $n \geq 2$ is a connected multi-graph embedded in the oriented $\bb{C}$ (resp. $\cp$) satisfying the following rules:
\begin{itemize}
\item The set of vertices $V$ is a finite or countable set and there is no accumulation point of vertices.
 \item The vertices are bipartite, $V=V_i \coprod V_o$: There is no edge connecting two vertices belonging to the same partition $V_i$ or $V_o$
 \item Every edge takes a value from $\lbrace 1,\dots, n \rbrace $. The set is given the natural cyclic order $1 <2 \dots n < 1$.
\item Every vertex is the endpoint of $n$ edges, one for each value.
\item The edges around a vertex belonging to $V_i$ have positive circular order.
      The edges around a vertex belonging to $V_o$ have negative circular order.
\item For any $k=1,\dots,n$, we define a map from $V_i$ to $V_o$ (and its inverse). $\mu_k(v)$ is the vertex adjacent to $v$ w.r.t. the edge $k$.
The composition  $\Sigma_k(v)=\mu_{k}\circ \mu_{k-1}(v)$ of two maps is a permutation of the inner (and of the outer) vertices. We suppose that for any $k$,
$\Sigma_k$ is not the identity.
\end{itemize}
We consider line complexes up to (orientation-preserving) homeomorphisms of $\bb{C}$ (resp. $\cp$).
\subsection{Rational functions}
Rational functions are conveniently represented by line complexes on $\bb{C}$.
In fact
\begin{lem}
For every abstract line complex on $\cp$ of index $n$, there exists a rational function $f$ (unique up to equivalence) with branching locus $B= \lbrace b_1,\dots,b_n\rbrace$
and that diagram.
\begin{proof}
This is essentially equivalent to the Riemann Existence Theorem. See \cite{elfving} for more details.
\end{proof}
\end{lem}

\subsection{Anharmonic Oscillators}
We are interested in studying the line complex of a function whose Schwarzian derivative is a polynomial (\ref{eq:schwarzian}).
Due to Section \ref{sec:nevanlinna}, such functions are defined by having
no critical points and $m+2$ direct transcendental singularities.

The branching locus of $f$ is the set of all asymptotic values, namely $\lbrace w_0,\dots,w_{m+1} \rbrace=\lbrace b_1,\dots,b_n \rbrace$.
The line complex of $f$ is an infinite diagram on $\bb{C}$ satisfying
\begin{itemize}
 \item[1]The index of the diagram is the cardinality of $\lbrace w_1,\dots,w_{m+1} \rbrace=\lbrace b_1,\dots,b_n \rbrace$
\item[2]For any permutation $\Sigma_{k}$, all non trivial cycles are infinite.
 \item[3] Let $\Sigma_{k}$ be the permutation corresponding to the asymptotic value $w_l=b_k$, then its cycle decomposition
has as many non-trivial cycles as Stokes sectors with asymptotic value $w_l$: the total number of non-trivial cycles is $m+2$.
\end{itemize}

\begin{figure}[htbp]
\input{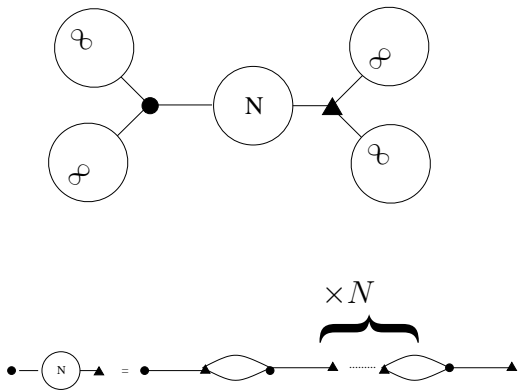}
 \caption{The line complex ${\cal{D}}^n$. We omit the labeling of the edges.}
\label{figure:harmonic}
\end{figure}

The following Theorem due to Nevanlinna establishes the bijection between line complexes with properties (1,2,3) and anharmonic oscillators.
\begin{thm}\label{thm:speisernevanlinna}
Given any abstract line complex of index $n$ on $\bb{C}$ satisfying (1,2,3) as above whose total number of non-trivial cycle is
$m+2$,
there exists a meromorphic function $f$, unique up to equivalence, with $m+2$ transcendental singularities
lying over $\lbrace b_1,\dots b_n \rbrace$, without critical points and with the given line complex.
\begin{proof}
\cite{nevanlinna32}.
\end{proof}
\end{thm}
\paragraph{An example}
As an example, we classify the line complexes corresponding to harmonic oscillators $y''(z)=(z^2-E)y(z)$ with just three distinct
asymptotic values - they are those line complexes of order $3$ with four non-trivial cycles, all of them infinite.
\begin{lem}\label{lem:harmspeiser}
The diagrams ${\cal{D}}^n, n \in \bb{N}$ represented in Figure \ref{figure:harmonic}, are all the
line complexes of harmonic oscillators $y''(z)=(z^2-E)y(z)$ with just three distinct
asymptotic values.
\end{lem}

\subsection{Rational approximations to Anharmonic oscillators}\label{sec:approx}
We have seen that branched covering  of the sphere with a finite number of transcendental singularities and no critical points -i.e.
anharmonic oscillators- are in bijection with line complexes satisfying properties (1,2,3) of Theorem \ref{thm:speisernevanlinna}.

The problem we address here is how to construct a meromorphic function $f$ given its line complex. In light of what we have shown so far
this is equivalent- if we restrict to cubic oscillators- to computing the location of poles of solutions of Painleve-I.

A solution of this problem was given by Nevanlinna \cite{nevanlinna32}. The idea is, given the line complex ${\cal{D}}$ of $f$, to construct a
sequence of rational function $f_k$ whose line complex ${\cal{D}}_k$ approximates ${\cal{D}}$.
As was already said we suppose that the branching locus is fixed and a curve $\gamma$ passing through it. Also we choose two points $P_i$
and $P_o$ belonging to the inner and outer circles cut by the Jordan curve.

The construction goes as follows:
\begin{itemize}
 \item[(i)] Choose a a vertex $v_0$ of ${\cal{D}}$ and finite portion $B_0$ of ${\cal{D}}$, to which $v_0$ belongs.
\item[(ii)] Similarly let $B_k$ the portion of $\cal{D}$ whose vertices and edges have distance not more than $k$ from $B_0$.
\item[(iii)]Embed $B_k$ in a disk. 
\item[(iv)] Reflect $B_k$ on the circle and switch the bipartition of the vertices to obtain a graph $\overline{B_k}$ with the correct
orientation of the edges.
\item[(v)] Finally, on $\cp$ draw the figure $B_k + \overline{B_k}$, by connecting every vertex in the frontier of $B_k$ with the corresponding
vertex in $\overline{B_k}$.
\end{itemize}
The resulting figure is a line complex ${\cal{D}}_k$ on $\cp$ whose Critical graph has a marked polygon, the one
to which $v_0$ belongs.

Let $f_k$ be the rational function with line complex ${\cal{D}}_k$
fixed by requiring: $0$ belong to the marked polygon and i)$f_k(0)=P_i$, ii)$ f_k'(0)=1$, while $\infty$ is an arbitrary point belonging to
mirror image of the marked polygon, for example $f_k(\infty)=P_o$.
\begin{thm}\label{thm:nevconvergence}
The sequence $f_k$ constructed above converges, uniformly with respect to the spherical metric on compact subsets of $\bb{C}$,
to the function $f$ with line complex ${\cal{D}}$
\begin{proof}
 \cite{elfving}.
\end{proof}
\end{thm}
As an example we construct here explicitly a sequence ${\cal{D}}^n_k, k \in \bb{N}$, related to the line complex ${\cal{D}}^n$
of the special harmonic oscillator constructed in Lemma \ref{lem:harmspeiser} and Figure \ref{figure:harmonic}.
They are shown in Figure \ref{figure:kharmonic} below.

\section{The Harmonic oscillator}
To illustrate his theory, Nevanlinna constructed explicitly \cite{nevanlinna30} -with ingenuity and owing to the big $\bb{Z}_{m+2}$ symmetry of the problem-
the sequence of rational approximations
of anharmonic oscillators whose potential is a monomial
\begin{equation*}
\lbrace f,z\rbrace= -2 z^m, m \in \bb{N} \; .
\end{equation*}
For the same purpose, we go a step further. Recall from Section \ref{sec:nevanlinna} that a meromorphic function satisfying
\begin{eqnarray*}
 \lbrace f,z\rbrace= -2 (z^2-E) 
\end{eqnarray*}
has four (counting multiplicities) asymptotic values $w_0,w_1,w_2,w_{-1}$, at least three of them distinct.
Here we concentrate on those $f$ such that two asymptotic values coincide, namely
\begin{equation}\label{eq:harmonicproblem}
w_0 \equiv \lim_{z \to + \infty}f(z)= w_2 \equiv \lim_{z \to -\infty}f(z)
\end{equation}
Latter condition is equivalent to the eigenvalue problem for the harmonic oscillator:
\begin{eqnarray*}
 y''(z)= (z^2-E) y(z) \mbox{ et } \\ \lim_{z \to + \infty}y(z)= \lim_{z \to -\infty}y(z)=0 \, .
\end{eqnarray*}
From standard Quantum Mechanics we know that this problem has a solution if and only if the
'energies' are quantized according to the rule $E=2n+1, n \in \bb{N}$.

We choose $w_0=w_2=1, w_1=i, w_{-1}=i$ and call $f_n(z)$ the solution of (\ref{eq:harmonicproblem});
we also choose the unit circle to be the Jordan curve needed to construct the line complex. 

With such a choice, the line complex of $f_n$ is the diagram ${\cal{D}}_n$ that was constructed in
Lemma \ref{lem:harmspeiser} and Figure \ref{figure:harmonic} \footnote{This follows from
Sturm-Liouville theory as the reader can verify.}.
Following the general procedure (Theorem \ref{thm:nevconvergence}),
for any diagram ${\cal{D}}_n$ we build a sequence of approximating diagrams ${\cal{D}}_{n\,k}, k \in \bb{N}$
on $\cp$. These are shown in Figure \ref{figure:kharmonic}.
\begin{figure}[htbp]
\input{betterkharmonic.pstex_t}
 \caption{The line complex ${\cal{D}}^n_k$}
\label{figure:kharmonic}
\end{figure}
\begin{lem}\label{prob:fnk}
Let $f_{n\,k}$ denote the (equivalence class of) rational function with line complex ${\cal{D}}_{n\,k}$.
Then $f_{n\,k}$ is a function of degree $2n+4k+2$ whose critical data are as follows
\begin{itemize}
 \item[(i)] $f_{n,k}$ have 4 critical points, $z_0,z_1,z_2,z_{-1}$ and three distinct critical values $b_0=1,b_1=i,b_{-1}=-i$.
\item[(ii)] There are two critical points $z_0$ and $z_2$ with the same multiplicity $2k$ and the same critical value $b_0$. And two
critical points  $z_1,z_{-1}$ with multiplicity $2n+2k+1$ and critical values $b_1$ and $b_{-1}$.
\item[(iii)] If we choose $z_0=1$, $z_1=i$, $z_{-1}=-i$ then $z_2=-1$ and $f_{n\,k}(1/z)=1/f_{n\,k}(z)=f_{n\,k}(-z)$
\end{itemize}
\begin{proof}
Points (i,ii) are evident from the line complex ${\cal{D}}_{n\,k}$.
To prove (iii), we fix  $z_0=1$, $z_1=i$, $z_{-1}=-i$ and notice that
 $f_{n\,k}(1/z)=1/f_{n\,k}(z)=f_{n\,k}(-z)$, due to the symmetry of the line complex. Therefore $z_2=-1$.
\end{proof}
\end{lem}
Functions $f_{n\,k}$ have just three distinct critical values and therefore are called \textit{Belyi functions} \cite{lando2004} (more about them in Section \ref{sec:riemann}).
They can be constructed recursively using a -partially conjectural- recursion formula.
First we tackle the case $n=0$.
\begin{thm}\label{thm:f1k}
Let us normalize $f_{0\,k}$ by choosing
$z_0=1,z_1=i,z_2=-1,z_{-1}=-i$ and let $f_{0\,k}=\frac{P_{0\,k}}{Q_{0\,k}}$, $P_{0\,k},Q_{0\,k}$ without common factors.
Then we can normalize
$P_{0\,k},Q_{0\,k}$ in such a way that they satisfy the following differential relations
\begin{eqnarray}\nonumber 
 P_{0\,k}''(z) &=&V_k(z)P_{0\,k-1}(z) \,, \; Q_{0\,k}''(z)=V_k(z)Q_{0\,k-1}(z)\,, \\
\label{eq:0diffrelation}  V_k(z) &=& (4k+2) (z^2(4k+1)-1) \; .
\end{eqnarray}
\begin{eqnarray} \nonumber 
P_{0\,k}(0)=- Q_{0\,k}(0)=1 &,& \,P_{0\,k}'(0)=Q_{0\,k}'(0)=-\gamma_k , \; \\ \label{eq:cauchykharmo}
 \gamma_k &=& 2 \prod_{l=1}^k\frac{4l+2}{4l} \; .
\end{eqnarray}
And hence
\begin{equation*}
 \frac{Q_{0\,k}''(z)}{Q_{0\,k}''(z)}= \frac{P_{0\,k-1}}{Q_{0\,k-1}}  \; .
\end{equation*}
Moreover, as $k \to +\infty$, $f_{0\,k}(\frac{x}{2\sqrt{k}})$ converges uniformly on compact subset of $\bb{C}$
to $f_0(x)$ - i.e. the solution of
\begin{equation}\label{eq:1schr}
\lbrace f_0, x \rbrace= -2 (x^{2}-1) \, , \;  w_0=w_2=1, w_1=i,w_{-1}=-i \; . 
\end{equation}
\begin{proof}
For sake of convenience we define $g_{k}=\frac{p_{k}}{q_{k}}=\frac{f_{0\,k}-1}{f_{0\,k}+1}$.
Since $P_{0\,k}=q_k-p_k$ and $Q_{0\,k}=q_k+p_k$ relations (\ref{eq:0diffrelation}) holds for $p,q$ if and only if they hold
for $P,Q$. The only effect of the transformation is to map the critical value $b_0=1 \to 0$.

As $f_{0\,k}$ (Lemma \ref{prob:fnk} (iii)), also the function $g_k$ has some useful symmetries, namely
$g_k(-z)=g_k(\frac{1}{z})=-g_k(z)$.

Since $g_k$ has degree $4k+2$ and $z=\pm 1$ is a zero of order $2k+1$, then
$p_k=c (z^2-1)^{2k+1}$ for some $c \in \bb{C}^*$. We choose $c=1$.
Due to the symmetry $f^1_k(-z)=-f^1_k(z)$, $q_{k}$ is an odd polynomial, $\deg{q_{k}}\leq 4k+1$.

By an explicit computation we verify that (\ref{eq:0diffrelation}) holds for $p_{k}$.
We prove that the same identities hold for the polynomial $q_{k}$.
To this aim we consider the polynomial $R_k(z)=q_{k}''(z)-V_k(z)q_{k-1}(z)$, whose degree is at most $4k-1$. 
Due to Lemma \ref{prob:fnk} (iii), 
\begin{equation}\label{eq:relproof}
p_k(z)\mp i q_k(z)=\pm i+ O((z\mp i)^{2k+2}) \forall k \geq 0.
\end{equation}
Due to the relation $p_k''=V_k(z)p_{k-1} $ then (\ref{eq:relproof}) implies that
$z=\pm i$ is a zero of order $2k$ for $R_k(z)$. Hence $R_k(z)$ vanishes identically.

We prove now (\ref{eq:cauchykharmo}). We notice that it is equivalent to
$p_k(0)=-1,p_k'(0)=0,q_k(0)=0,q_k'(0)=\gamma_k$. Since the first three relations are already proven, we must verify
the latter. Let $q_k(z)=a_k z^{4k+1} + O((z^{4k-1})$.
Due to (\ref{eq:0diffrelation}) then
$\frac{a_k}{a_{k-1}}=\frac{4k+2}{4k}$. Hence $ a_{k}= a_0 \prod_{l=1}^k\frac{4l+2}{4l}$.
Remarkably, $q_k'(0)=a_k$. In fact,
due to the symmetry $g_k(1/z)=-g_k(z)$ and since $z^{4k+2} p_k(1/z)=-p_k(z)$,
we have that $q_k(z)=z^{4k+2}q_k(1/z) \Rightarrow q_k'(0)=a_k$.
By a direct computation we verify that $g_0(z)=\frac{z^2-1}{2z} \Rightarrow a_0=2$. Relation (\ref{eq:cauchykharmo}) is proven.

We note that due to
(\ref{eq:0diffrelation}) if, under the scaling $x=2\sqrt{k}z$, $p_{k}$ and $q_{k}$ converge to some entire non-zero functions,
then their limits must satisfy $y''(x)=(x^2-1)y(x)$. Trivially the polynomials
$p_{k}$ converge to $-e^{-z^{2}}$, which is indeed a solution of the Schr\"odinger equation.
Since $q_k(0)=0, q'_{k}(0)= \gamma_k = \frac{4 \Gamma(3/2+k)}{\sqrt{\pi}\Gamma(k+1)}$, we have that $q_{k}(0) \to 0, q'_{k}(0)
\to \frac{2}{\sqrt{\pi}}$. Hence if $q-k$ converges it converges to a non-zero functions.

We can prove its convergence using (\ref{eq:0diffrelation}). First we change variable and choose $s_k(t)=-i q_k(i t/(2\sqrt{k}))$ so that
in (\ref{eq:0diffrelation}) $V_k$ becomes $(4k+2) (z^2(4k+1)+1)$. We let $s_k(t)=\sum_{l=0}^{2k} b^k_l t^{2l+1}$. Using (\ref{eq:0diffrelation})
and $s_0=2t$, we prove by (a double) induction that there exist $C_l, l \in \bb{N}$ such that
\begin{eqnarray*}
0\leq b_l^{k-1} \leq b_l^k \leq C_l \, , \;  \forall l,k \\
C_{l+1} \leq \frac{1}{2l+1} (C_l+C_{l-1}) \, ,  \; \forall l \; .
\end{eqnarray*}
The first inequality proves that $b_l^k$ converges to some positive value $b_l^k$ while the second shows that $s_k$ converges to some function with an infinite radius of
convergence. Hence $g_k$ converges in the scaling limit to a solution of (\ref{eq:1schr}). We let the reader verify that  $w_0=w_2=0$, $w_1=i, w_{-1}=i$.
\end{proof}
\end{thm}
We leave the generalization of Theorem \ref{thm:f1k} for the general $f_{n\,k}$ as a conjecture.
\begin{conj}\label{conj:fnk}
Let us normalize $f_{n,k}$ by choosing
$z_0=1,z_1=i,z_2=-1,z_{-1}=-i$ and let $f_{n,k}=\frac{P_{n\,k}}{Q_{n\,k}}$, $P_{n\,k},Q_{n\,k}$ without common factors.
Then we can normalize
$P_{n\,k},Q_{n\,k}$ in such a way that they satisfy the following differential relations
\begin{eqnarray}\nonumber
 P_{n\,k}''(z)&=&V_{n\,k}(z)P_{n\,k-1}(z) \,, \; Q_{n\,k}''(z)=V_{n\,k}(z)Q_{n\,k-1}(z)\,, \\
\label{eq:partialschr}  V_{n\,k}(z) &=& (4k+2n+2) (z^2(4k+2n+1)-2n-1) .
\end{eqnarray}
\begin{eqnarray} \nonumber
P_{n\,k}(0)=(-1)^{n+1} Q_{n\,k}(0)=1 &,& \,P_{n\,k}'(0)=(-1)^nQ_{0\,k}'(0)=-\gamma_{n\,k} \; , \\
\label{eq:ncauchykharmo}  \gamma_{n\,k} &=& (2n+2) \prod_{l=1}^k\frac{4l+2n+2}{4l} \; .
\end{eqnarray}
And hence
\begin{equation*}
 \frac{Q_{n\,k}''(z)}{Q_{n\, k}''(z)}= \frac{P_{n\,k-1}}{Q_{n\,k-1}}  \; .
\end{equation*}
Moreover, as $k \to +\infty$, $f_{n\,k}(\frac{x}{2\sqrt{k}})$ converges uniformly on compact subset of $\bb{C}$
to $f_n(x)$ - i.e. the solution of
\begin{equation}\label{eq:nschr}
\lbrace f_{n}, x \rbrace= -2 (x^{2}-2n-1) \, , \;  w_0=w_2=1, w_1=i,w_{-1}=-i \; . 
\end{equation}
\end{conj}
Notice that (\ref{eq:ncauchykharmo}, \ref{eq:nschr}) can be proven, provided (\ref{eq:partialschr}) holds,
using essentially the same proof of Theorem \ref{thm:f1k}.
Relations (\ref{eq:ncauchykharmo}) are rather interesting because they solve what, in the large $k$ limit,
is the central connection problem for the harmonic oscillator.
\begin{rem*}
We stress that Lemma \ref{prob:fnk}(i,ii,iii,iv) allow us to compute exactly the functions $f_{n\,k}$, without invoking
the differential relations (\ref{eq:0diffrelation}, \ref{eq:partialschr}). How this is possible will be shown in the next Section
of the paper, see Theorem \ref{thm:criticalmap}.
\end{rem*}
Here we list few of the functions $f_{n\,k}(z)$ and we plot the corresponding Critical Graph, which is
the the inverse image of $|f_{n\,k}|=1$. In the images below, in purple
is $|f_{n\,k}(z)| <1$, in grey is $|f_{n \,k}(z)| >1$. We let the reader associate to any image, the corresponding function.
\begin{itemize}
 \item $f_{0\,1}(z)=\frac {1 - 3 z - 3 z^2 + 2 z^3 + 3 z^4 - 3 z^5 - z^6} {-1 - 3 z + 
   3 z^2 + 2 z^3 - 3 z^4 - 3 z^5 + z^6}$
\item $f_{0\,2}(z)= \frac{4-15 z-20 z^2+20 z^3+40 z^4-58 z^5-40 z^6+20 z^7+20 z^8-15 z^9-4 z^{10}}
{-4-15 z+20 z^2+20 z^3-40 z^4-58 z^5+40 z^6+20 z^7-20 z^8-15 z^9+4 z^{10}}$
\item $f_{1\,1}(z)=\frac{3+16 z-36 z^2-48 z^3+50 z^4+48 z^5-36 z^6-16 z^7+3 z^8}{3-16 z-36 z^2+48 z^3+50 z^4-48 z^5-36 z^6+16 z^7+3 z^8} $
\item $f_{2\,1}(z)= \frac{2 - 15 z - 50 z^2 + 100 z^3 + 140 z^4 - 154 z^5 - 140 z^6 + 100 z^7 + 
 50 z^8 - 15 z^9 - 2 z^{10}}{-2 - 15 z + 50 z^2 + 100 z^3 - 140 z^4 - 154 z^5 + 140 z^6 + 
 100 z^7 - 50 z^8 - 15 z^9 + 2 z^{10}}$
\end{itemize}

\begin{minipage}[c]{.40\textwidth}
\begin{center}
\includegraphics[width=4cm]{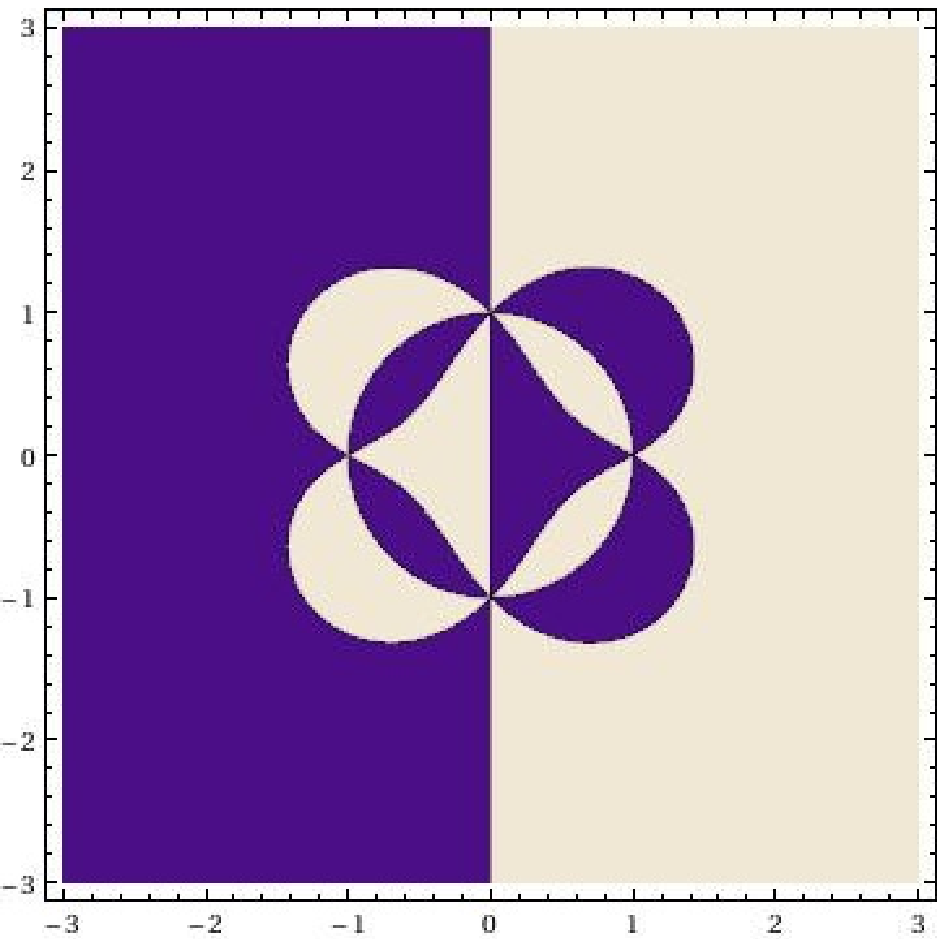}
\end{center}
\end{minipage} \hspace{1cm}
\begin{minipage}[c]{.40\textwidth}
\begin{center}
\includegraphics[width=4cm]{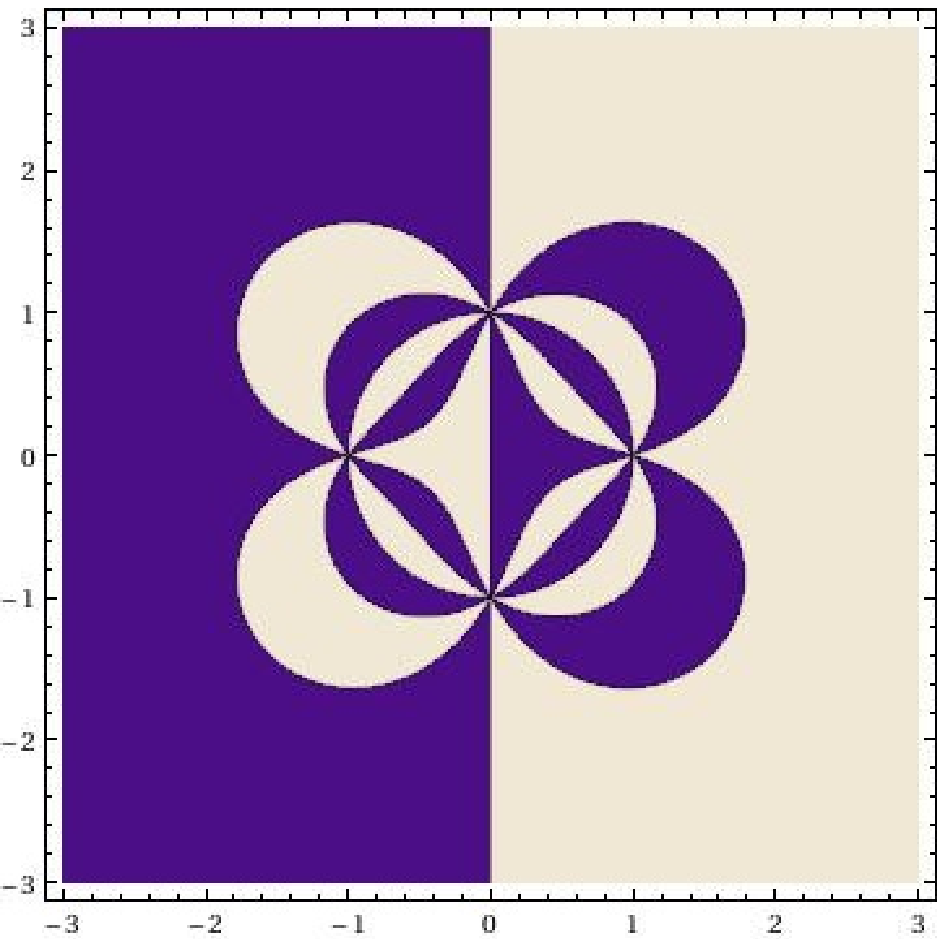}
\end{center}
\end{minipage}

\begin{minipage}[c]{.40\textwidth}
\begin{center}
\includegraphics[width=4cm]{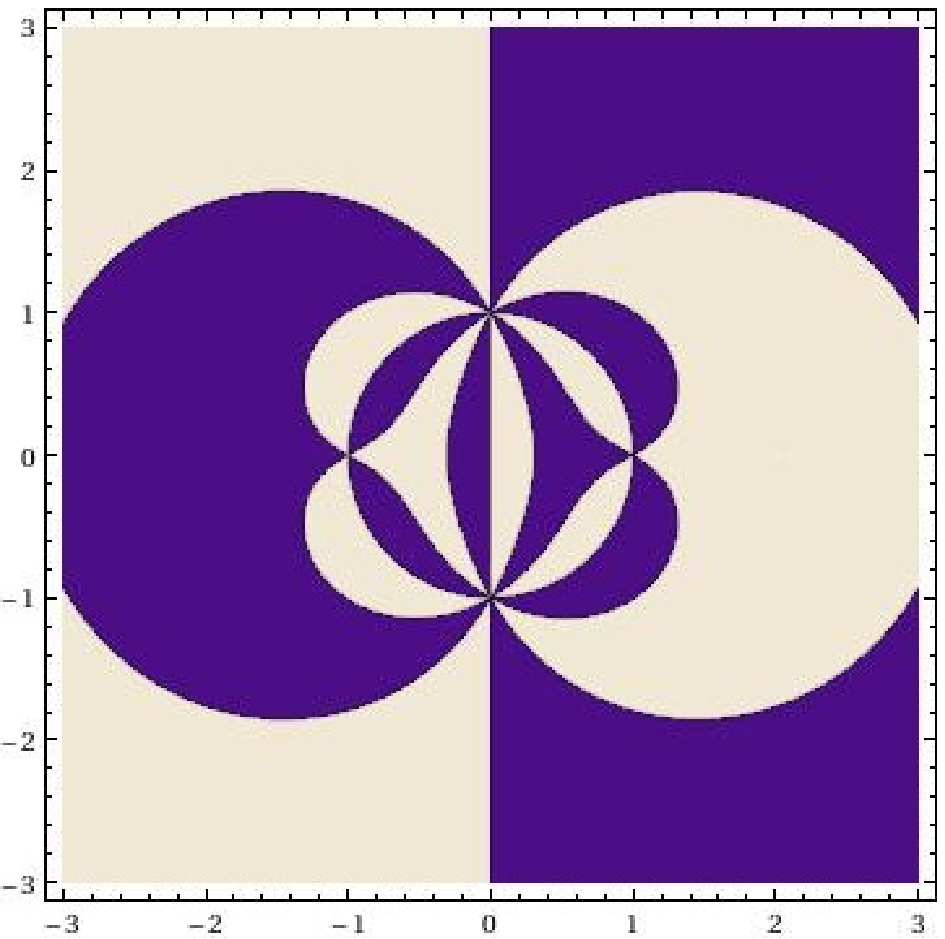}
\end{center}
\end{minipage} \hspace{1cm}
\begin{minipage}[c]{.40\textwidth}
\begin{center}
\includegraphics[width=4cm]{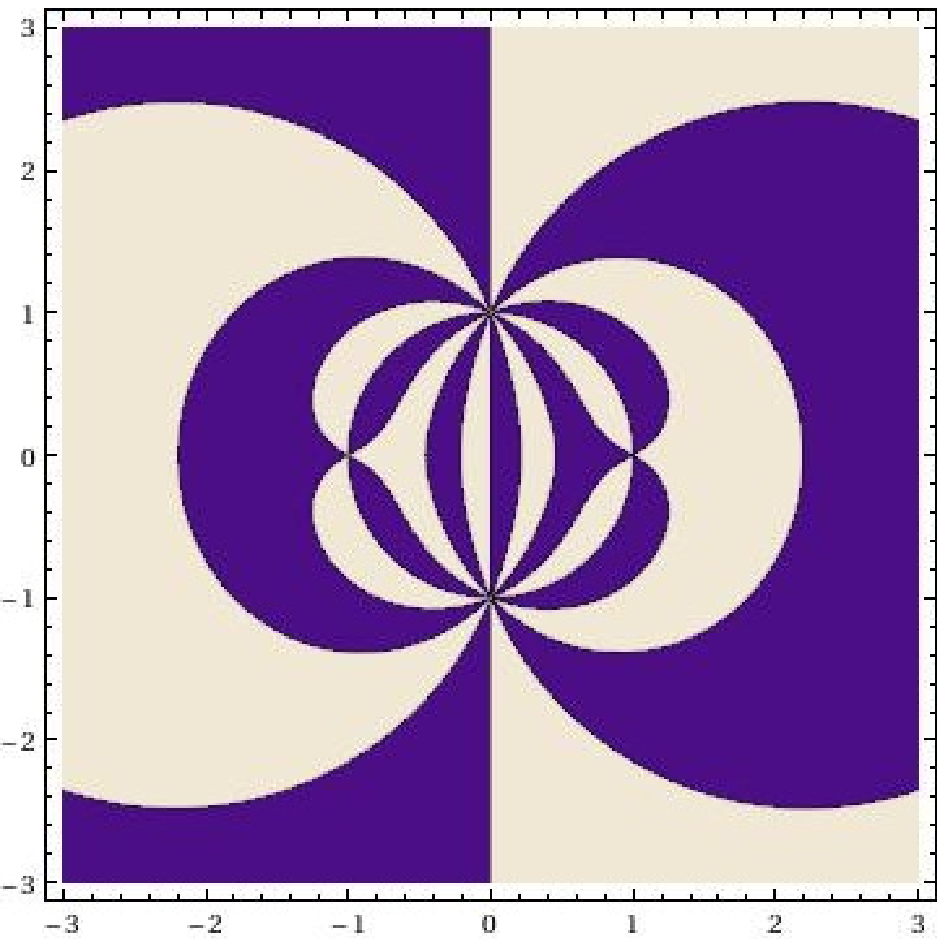}
\end{center}
\end{minipage}

\section{The Riemann problem}\label{sec:riemann}
In the previous Sections, we have shown that computing the location of poles of Painleve I transcendents
is equivalent to computing some infinitely-sheeted branched coverings of the sphere. These in turn can be constructed as limit of a sequence
of coverings of the sphere with a finite number of sheets. In other words, we have reduced the original problem about Painleve transcendents
to the construction of a rational function once its critical values and its monodromy representation are given.
We name this latter problem the Riemann problem following Plemelj's nomenclature. Plemelj solved the Riemann problem
by what is now called Riemann-Hilbert technique \cite{plemelj}. We give here another treatment, which is algebraic in nature.
 
To start with, we introduce a simplification of the Riemann problem, that may admit more than one solution, but that can
be effectively solved by purely algebraic means, as we show below.
\begin{Def}
We call Hurwitz problem the problem of constructing a rational function
once the following data are given: the critical values,
the number of corresponding critical points and their multiplicities.
\end{Def}
In general - if the number of critical points is larger than three - a Hurwitz problem do not uniquely determine
a single equivalence class of rational functions but a discrete family of them, because a finite set of
inequivalent monodromy representations may share the same cycles structure \footnote{A Hurwitz problem have no solutions at all,
if the data of the problem do not correspond to the cycles structure of any monodromy representation}.

Let us discuss the solution of the Hurwitz problem.
\begin{Def}
Let $f$ be a rational function, we call critical data the collection of all the critical points of $f$, their multiplicities and their critical values.
\end{Def}
Our method of solution of the Hurwitz problem is based on the following observation: If we represent
the rational function $f(z)=\frac{P(z)}{Q(z)}$ as a ratio of two polynomials, then
the critical data impose some linear relation on $P$ and $Q$, which can be solved to reconstruct $f$.
In fact
\begin{lem}\label{lem:point}
Given $P,Q$ polynomials with no common factors, let $f=\frac{P}{Q}, \deg f= n$ be their ratio.
Then

$z^* \neq \infty $ is a critical point of multiplicity $v$ with critical value $b^* \neq \infty$ if and only if
\begin{eqnarray}\label{eq:crit}
P^{(l)}(z^*) &=& b^*Q^{(l)}(z^*) \, , \; l=0,\dots,v \; \\ \nonumber
P^{(v+1)}(z^*)&\neq& b^*Q^{(v+1)}(z^*)
\end{eqnarray}
Here $P^{(l)},Q^{(l)}$ denote the l-th derivative of $P$ and $Q$.

$z^* \neq \infty $ is a critical point of multiplicity $v$ with critical value $b^* = \infty$ if and only if
\begin{eqnarray}\label{eq:crit0}
Q^{(l)}(z^*) &=& 0 \, , \; l=0,\dots,v \; \\ \nonumber
Q^{(v+1)}(z^*)&\neq& 0
\end{eqnarray}

Let $p(t)=t^nP(\frac{1}{t}),q(t)=t^nQ(\frac{1}{t})$.
$b^* = \infty $ is a critical point of multiplicity $v$ with critical value $b^* \neq \infty$ if and only if
\begin{eqnarray}\label{eq:infcrit}
p^{(l)}(0) &=& b^*q^{(l)}(0) \, , \; l=0,\dots,v \; \\ \nonumber
p^{(v+1)}(0)&\neq& b^*q^{(v+1)}(0)
\end{eqnarray}

$b^* =\infty $ is a critical point of multiplicity $v$ with critical value $b^* = \infty$ if and only if
\begin{eqnarray}\label{eq:infcrit0}
q^{(l)}(z^*) &=& 0 \, , \; l=0,\dots,v \; \\ \nonumber
q^{(v+1)}(z^*)&\neq& 0
\end{eqnarray}

\end{lem}
The reader should notice that equations (\ref{eq:crit}, \ref{eq:crit0}, \ref{eq:infcrit}, \ref{eq:infcrit0}) are linear
in the space of ordered pairs of polynomials $\bb{C}[X] \oplus \bb{C}[X]$.

Motivated by the previous Lemma we define a linear operator on the space of pairs of polynomials
\begin{Def}\label{def:criticalmap}
Given a triplet $(z,\nu,b)\in\cp \times \bb{N} \times \cp$ we define a linear operator $A_n(z,\nu,b)$ from the space of
pairs of polynomials of degree at most $n$ to $\bb{C}^{\nu+1}$:
\begin{equation*}
 A_n(z,\nu,b): \bb{C}[X]_n \oplus \bb{C}[X]_n \to \bb{C}^{\nu+1} \; .
\end{equation*}
We call $A$ the \textbf{critical evaluation map} and define its action as follows:
\begin{itemize}
 \item If $z\neq \infty$ and $b \neq \infty$ then $A(P,Q)= \oplus_{l=0}^{\nu}P^{(l)}(z) - w Q^{(l)}(z) $.
\item If $z\neq \infty$ and $b = 0 $ then $A(P,Q)= \oplus_{l=0}^{\nu} (- Q^{(l)}(z) ) \, $.
\item If $z = \infty $ and   $b \neq \infty$, then $A(P,Q)= \oplus_{l=0}^{\nu} p^{(l)}(0) - w q^{(l)}(0)$, where
$p(t)=t^nP(\frac{1}{t}),q(t)=t^nQ(\frac{1}{t})$.
\item If $z = \infty $ and   $b = \infty$, then $A(P,Q)= \oplus_{l=0}^{\nu} (- q^{(l)}(0)) $, where
$p(t)=t^nP(\frac{1}{t}),q(t)=t^nQ(\frac{1}{t})$.
\end{itemize}
More generally, we extend the critical evaluation map to sequences of triplets $\lbrace(z_i,\nu_i,w_i)\rbrace_{i=1,\dots,s}$:
it is the direct sum of the critical evaluation maps $\oplus_i A_n(z_i,\nu_i,w_i)$
\end{Def}

Due to Lemma \ref{lem:point}, if
a sequence $\lbrace(z_i,\nu_i,b_i)\rbrace$ represents the critical data
of a rational function $f=P/Q$ then $(P,Q)$ belongs to the kernel of the critical evaluation map $A(\lbrace(z_i,\nu_i,b_i)\rbrace)$.

We show that the converse is true in Theorem \ref{thm:criticalmap} below.
\begin{thm}\label{thm:criticalmap}
Given a sequence $\lbrace(z_i,\nu_i,b_i)\rbrace_{i=1}^{s}$ such that
\begin{itemize}
\item[(i)] $\sharp \lbrace b_1, \dots,b_s\rbrace \geq 3 $
\item[(ii)] $\sum_i\nu_i$ is even
\item[(iii)] for all $i$, $\sum_{b_j=b_i} \nu_j  \leq \sum_l\frac{\nu_l}{2}$
\item[(iv)] all the $z_i$'s are distinct
\end{itemize}
Let $n=1+\frac{\sum_i\nu_i}{2}$ and ${\cal{A}}=A_n(\lbrace(z_i,\nu_i,b_i)\rbrace_{i=1}^{s})$ be the critical evaluation map
of the sequence, acting on the space of pairs of polynomials of degree at most $n$.

Then
\begin{itemize}
 \item(a) If $(P,Q)$ is a non-trivial element of the kernel of ${\cal{A}}$, then
$(P,Q)$  have no common factors and the sequence $\lbrace(z_i,\nu_i,b_i)\rbrace_{i=1}^{s}$
is the critical data of $f=\frac{P}{Q}$.
\item(b) The kernel of ${\cal{A}}$ has at most one dimension.
\end{itemize}
\begin{proof}
To avoid considering particular sub-cases of the same proof and without loss of generality we may suppose $z_i\neq \infty, \forall i$.

a) Let $(P,Q)$ belong to the kernel of $\cal{A}$. Uniquely up to a norming constant, we can write 
$$(P(z),Q(z))=\prod_{i=1}^s(z-z_i)^{k_i}r(z)(p(z),q(z)) \, ,$$
where $k_i\leq\nu_i $, $r(z_i)\neq0, \forall i$ and $p,q$ have no common factors.

Suppose $p/q$ is not a constant function: neither $p$ nor $q$ are zero and the they are not both constant polynomials.
Then $f(z_i)=b_i$ and $z_i$ has multiplicity $\mu_i \geq  \nu_i-k_i \geq 0$.

Hence, due to Riemann-Hurwitz,
$$\deg{f} \geq 1+ \sum_i\frac{\mu_i}{2} \geq 1+\sum_i\frac{\nu_i-k_i}{2}= n-\frac{\sum_ik_i}{2} \; . $$
At the same  time $$\deg{f} \leq n-\sum_i k -\deg r \; ,$$ since $\deg P\leq n-\sum_ik_i -\deg r$ and $\deg Q \leq n-\sum k_i -\deg r$.

Hence, for both inequalities to be satisfied, it is necessary that
$k_i=0, \forall i$ and $r=\mbox{const}$. Hence  $(P,Q)$ have no common factors and $\lbrace(z_i,\nu_i,w_i)\rbrace_{i=1}^{s}$
represents the critical data of $f=\frac{P}{Q}$.

To conclude the proof we must show that $p/q$ is not a constant. We argue by contradiction.

Suppose $p/q$ is a constant; by a M\"obius transformation, we can always reduce to the case it is $0$:
We can suppose that $p=0$. Then $p=P=0$ and $Q$ satisfies $Q^{(j)}(z_i)=0, i=,0,\dots,\nu_i$ for all $i$ s.t. $b_i\neq 0$.
Hence $Q$ has at least $\sum_{b_i \neq 0} (\nu_i+1)$ zeroes.

By hypothesis (i,iii), the number of the zeroes is greater or equal
than $\frac{\sum_i \nu_i}{2}+2=n+1$. Hence $Q$ is zero. Hence $(P,Q)$ is the zero vector. Which is a contradiction.

ii)Let $(P,Q)$, $(p,q)$ two elements of the kernel of ${\cal{A}}_n$. To prove the thesis we must show that $f=P/Q$ is equal to $g=p/q$.

Suppose $f-g\neq 0$ then $z_i$ is a zero of order $\nu_i+1$ of $f-g$ for every $i=1,\dots,p$. Hence $f-g$ has at least $\sum_i\nu_i+s=2n+s-2$ zeroes,
counting multiplicities. However $\deg{f-g}\leq 2 n$ which is less than the number of zeros of $f-g$, by the hypothesis on $s$.
This is a contradiction.
\end{proof}
\end{thm}

\subsection{The solution of the Hurwitz problem}
\begin{Def}
Let $\lbrace (\nu_i, b_i) \rbrace_{i=1}^{s}$ be a sequence of multiplicities and branch points.
We call $\lbrace (\nu_i, b_i) \rbrace_{i=1}^{s}$ an admissible for a Hurwitz problem
provided
\begin{itemize}
\item[(i)] $\sharp \lbrace b_1, \dots,b_s\rbrace \geq 3 $.
\item[(ii)] $\sum_i\nu_i$ is even.
\item[(iii)] for all $i$, $ \sum_{b_j=b_i} \nu_j \leq \sum_l\frac{\nu_l}{2}$.
\end{itemize}
\end{Def}
Given admissible data $\lbrace (\nu_i, b_i) \rbrace_{i=1}^{s}$, we can solve the Hurwitz problem using the tools
developed so far, in particular Lemma \ref{lem:point} and Theorem \ref{thm:criticalmap} above.
In fact, they show that it is sufficient to find all the sequences $\lbrace (z_i,\nu_i,b_i)\rbrace_{i=1}^{p}$
such that the associated critical evaluation map -restricted to
pairs of polynomials of degree $n=1 + \frac{\sum_i\nu_i}{2}$ - has a non trivial kernel.

To this aim we start by normalizing the problem by fixing three of the $z_i$'s, say $z_1,z_2,z_3$ to be $0,1,\infty$ and we consider
the remaining critical points as the unknowns of the problem. 
 
Since critical evaluation map sends $$\bb{C}[X]_n \oplus \bb{C}[X]_n \cong \bb{C}^{2n+2} \to \bb{C}^{l}, l=\sum_i(\nu_i+1)=2n+s-2 \; ,$$
we have two cases, $s=3$ and $s>3$. In the first case, the critical evaluation map has certainly a non-trivial kernel and the unique solution
of the Hurwitz problem is readily solved by computing a vector of the kernel.

Conversely, if $s>3$, we fix a basis of $\bb{C}[X]_n \oplus \bb{C}[X]_n$ and let $v_1,\dots v_{2n+2}$
denote the columns of the critical evaluation map matrix - notice that $v_i$ are polynomial functions of $z_4,\dots,z_s$.
Then the critical evaluation map has a non trivial kernel if and only if
the following system of polynomial equations are satisfied:
\begin{equation}\label{eq:wedge}
\star \left( v_1 \wedge v_2 \dots \wedge v_{2n+2} \right) =0 \; , 
\end{equation}
$\star$, $\wedge$ being respectively the Hodge dual and exterior product.
\begin{Def}
We say that a solution of (\ref{eq:wedge}) is admissible if all the $z_j, j=1,\dots,s$ are pairwise distinct.
\end{Def}
\begin{lem}\label{lem:admissible}
Let a Hurwitz problem with admissible data $\lbrace (\nu_i, w_i) \rbrace_{i=1}^{s}$ be given. Then
the critical points of a function satisfying the Hurwitz problem 
are admissible solutions of (\ref{eq:wedge}).
Conversely for every admissible solution $z_1,\dots z_s$ of (\ref{eq:wedge}) the sequence  $\lbrace z_i, \nu_i, b_i\rbrace_{i=1,\dots,s}$
is the critical data of a unique function, and this function satisfies the Hurwitz problem.
\begin{proof}
It follows immediately from Lemma \ref{lem:point} and Theorem \ref{thm:criticalmap}.
\end{proof}
\end{lem}
Summing up, we have shown that the Hurwitz problem is essentially equivalent to the system of polynomials equation (\ref{eq:wedge}).
\begin{rem*}
A finite set of different admissible solutions of (\ref{eq:wedge}) may lead to the same solution of the Hurwitz problem,
if the latter has some symmetries. 
\end{rem*}
\begin{rem*}
Note that (\ref{eq:wedge}) is a system of $\binom{l}{2n+2}$ equations, where $l=\sum_i (\nu_i+1)=2n-2+k$
in $k-3$ unknowns. However, we know that the system has only a finite number of admissible solutions since
the Hurwitz problem has a finite number of solutions.
\end{rem*}
\paragraph{The solution of the Riemann problem} 
Given a Riemann problem, i.e. a branching locus and a monodromy representation, we associate to it a Hurwitz problem
by extracting from the monodromy representation its cycles structure. The solution of the Riemann problem
is one of the admissible solutions of
the associated Hurwitz problem.
To select the correct one it is sufficient to compute the monodromy - for example by plotting the Critical
Graph - of all solutions of the Hurwitz problem. However, it may happen that further insights (e.g. inequalities on the
location of the critical points) about the specific problem
allow to \textit{a-priori} select the correct solution without computing the monodromy. This is the case, for example, of
some Riemann problems associated with poles of the integrale tritronquee, see Section \ref{sec:polesriemann} below.

\paragraph{Belyi Functions}
We end our discussion of the Riemann problem by introducing the \textit{Belyi} functions.
They are defined to be those rational functions (or more generally, those meromorphic functions on a compact Riemann surface)
with just three critical values, namely $0,1,\infty$.

The importance of Belyi function lies in the following important property.
\begin{thm}\label{thm:belyi}
If a rational function $f$ has critical values $0,1,\infty$, then $f$ is equivalent to a function $g$
defined over a number field $K$, $\bb{Q}\subseteq K \subset \overline{\bb{Q}}$ (here $\overline{\bb{Q}}$ is the algebraic closure of $\bb{Q}$).

In other words,  $\forall x \in K \, , \; g(x) \in K$. Equivalently let $g=P/Q$, then we can choose $P=\sum a_n z^n, Q= \sum b_n z^n$
in such a way that $\forall n, \, a_n,b_n \in K$.
\begin{proof}
See \cite{belyi}. 
\end{proof}
\end{thm}
Our solution of the Riemann problem give a hint of why the Theorem holds. In fact, system (\ref{eq:wedge})
has integer coefficients so that we can expect its solutions to be algebraic numbers.
However, a full proof would require to identify the 0 dimensional component of solutions
of (\ref{eq:wedge}) as well as deep instruments from Elimination Theory, such as the \textit{Shape Lemma}. We will not go into that.
\section{Poles of the Tritronquee Solution}\label{sec:polesriemann}
According to Theorem \ref{thm:polesvalues}, poles of the Tritronquee solution are in correspondence with those
cubic polynomials $V(\lambda;a,b)=4\lambda^3-2a\lambda-28b$ such that the solution of the Schwarzian equation
has just three distinct asymptotic values, namely
\begin{eqnarray}\nonumber 
 \lbrace f,z\rbrace &=& -2 V(\lambda;a,b) \; , \\ \label{eq:cubicschw}
w_1=w_{-2} &,& w_{-1}=w_2  \; .
\end{eqnarray}
Our aim is to construct such transcendentals functions as limit of rational functions. Since
the number of distinct asymptotic values, three in this case, is the number of critical values of the rational approximants, then
all the rational functions that we are looking for are \textit{Belyi functions}.

Contrary to the case of the harmonic oscillator, we will be unable to give
explicit general formulas as we expect the approximating functions to be defined -generically-
over some number field different from $\bb{Q}$.

\begin{figure}[htbp]
\input{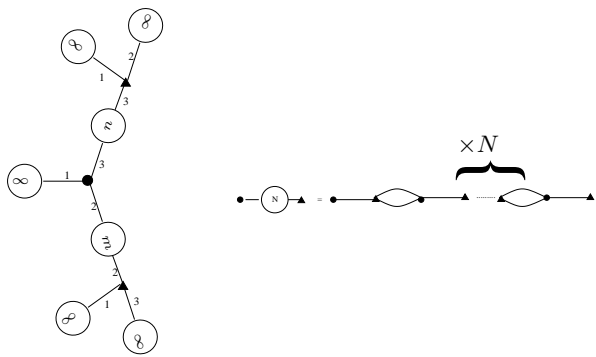}
 \caption{The line complex ${\cal{D}}^{n,m}$}
\label{figure:tritronquee}
\end{figure}

We proceed by classifying all line complexes corresponding to the poles of the tritronquee, hence by classifying poles
themselves.
\begin{lem}
The line complexes of cubic oscillators with only three distinct asymptotic values are classified by two integers
$m,n \in \bb{N} \times \bb{N}$. We let ${\cal{D}}^{n,m}$ denote them. They are represented in Figure \ref{figure:tritronquee}. 
\end{lem}

\subsection{Lines complexes and WKB approximation}\label{subsec:wkb}
Before tackling the construction of the rational approximants, we answer here
the following question: where is, approximately, the pole of the tritronquee solution
corresponding to the line complex ${\cal{D}}^{n,m}$? To this aim we have to analyze the connection between the line complexes
and the WKB analysis of solutions of the cubic oscillators. Here we simply state this connection without entering into the details of the proof,
as it would require the introduction of heavy machinery from WKB analysis of our previous works.
\textit{In this Subsection we suppose the reader
to be be familiar with the results of \cite{piwkb}, \cite{piwkb2}}.

In \cite{piwkb} we have derived a system of two equations (equation (25) of \cite{piwkb}),
called Bohr-Sommerfeld-Boutroux system, for the unknowns $a,b$ coefficients of
the cubic potential, describing the asymptotic location of poles of the tritronquee solution.
Solutions of that systems are discrete and classified by two positive integers $l,k$.

Given two quantum numbers $l,k$ and the unique solution $\alpha^{l,k},\beta^{l,k}$ of the Bohr-Sommerfeld-Boutroux equations,
for $l,k$ big enough there exists a unique pole $a$ of the tritronquee close (in some precise sense) to $\alpha^{l,k}$, see Theorem 1 in \cite{piwkb2}.

The answer to the above question is as follows: the pole of the tritronquee solution
corresponding to the line complex ${\cal{D}}^{n,m}$ is, at least for large $n,m$, the pole close to the solution of the Bohr-Sommerfeld-Boutroux system
with indices $l=n+1,k=m+1$.

Let us give some hints of the proof.
Let us normalize $f^{n,m}$ by choosing $w_0=0,w_1=1,w_2=\infty$: then the Critical Graph is the union of level curves $Re{f^{n,m}}=0$ and
the numbers $n$, $m$ count the distinct curves $Re{f^{n,m}}=0$ that asymptotically lie in the Stokes sectors $S_0$ and
$S_2$, $S_{-2}$ (Stokes sectors as defined in (\ref{eq:cubicsector})).

The WKB approximation of the cubic oscillator \cite{piwkb}, allow us to approximate the solution $f$ of $\lbrace f,z\rbrace = -2 V(z;a,b)$
(fixed by choosing $w_0=0,w_1=1,w_2=\infty$), as $f(z) \sim \tilde{f}(z) =  e^{S(z)}, S(z)=\int^z\sqrt{V(u;a,b)} du$, where $\sqrt{V}$
is defined in the plane with some cuts.
Clearly given a representation of $\tilde{f}$ as above, the lines $Re \tilde{f}=0$ correspond to $Im S= d \pi, d \in \bb{Z}$.

Remarkably, the indices $l,k$ of the Bohr-Sommerfeld-Boutroux system count the number of lines $Im S= d \pi$ that connect $S_0$ with $S_2$, $S_{-2}$
\footnote{Notice that the Stokes graph of solution of the Bohr-Sommerfeld-Boutroux system (Figure 4 in \cite{piwkb})
is the bifurcation diagram of the lines $Re S=0$.}.

Hence we have two different counts for the number of of lines $Re{f}=0$ connecting $S_0$ with $S_2$ and $S_{-2}$. We have the theoretical (exact)
one given by the line complex, and the approximate one given by the WKB analysis. For $n,m$ big enough, the two counts can be proven to coincide.

\subsection{Rational Approximants}
In the present Section, we start studying rational approximation to \ref{eq:cubicschw}. As it as already remarked, this is a preliminary study.

To construct rational approximants to the function with lines complex ${\cal{D}}^{n,m}$, we follow the general procedure
(see Theorem \ref{thm:nevconvergence}) and
for any diagram ${\cal{D}}^{n,m}$ we build a sequence of approximating diagrams ${\cal{D}}^{n,m}_k, k \in \bb{N}$
on $\cp$. These are shown in Figure \ref{figure:ktritronquee}.
\begin{figure}[htbp]
\input{ktritronquee.pstex_t}
 \caption{The line complex ${\cal{D}}^{n,m}_k$}
\label{figure:ktritronquee}
\end{figure}
We let $f_k^{n,m}$ denote the (equivalence class of) rational function(s) whose line complex is ${\cal{D}}^{n,m}_k$.
\begin{lem}\label{lem:ktritronquee}
The function $f_k^{n,m}$ has five critical points $z_0,z_1,z_2,z_{-1},z_{-2}$, three distinct critical values
$b_0,b_1=b_{-2},b_{-2}=b_{-1}$ and multiplicities
\begin{eqnarray*}
 \nu_0=2(m+n+k+1), \nu_1 \!\!&=&\!\!2k, \nu_2=2(n+k)+1 \, ,\\
 \nu_{-1} \!\!&=&\!\! 2k, \nu_{-2}=2(m+k)+1 \; .
\end{eqnarray*}
Let us fix $f_k^{n,m}$ by choosing $b_0=z_0=\infty, b_2=z_2=0, b_{-2}=z_{-2}=1$. Then
\begin{itemize}
 \item[(i)]$\overline{f^{n,m}_k(\overline{z})}=f_k^{n,m}(z)$, where 
$\overline{~~~}$ stands for complex conjugation.
\item[(ii)] $f_k^{n,m}(-z+1)=-f_k^{m,n}(z)+1$. Hence, if we denote 
by $z_{\pm 1}(m,n)$ be the critical points $z_{\pm1}$ of the functions $f^{n,m}$, we have
\begin{equation}\label{eq:z+z}
z_{1}(m,n)=-z_{-1}(n,m) +1   \; .
\end{equation}
 \item[(iii)]Fix $w_0=z_0=\infty, w_2=z_2=0, w_{-2}=z_{-2}=1$ then
\begin{equation}\label{ineq:zz}
z_1,z_{-1} \in \bb{R} \, , \; z_{-1}>1 \mbox{ et } z_1<0 \; . 
\end{equation}
\end{itemize}
\begin{proof}
The multiplicities of the critical points as well as the symmetries are read directly from the line complex.

Here we prove (i) and leave (ii) to the reader. The critical graph of $f^{n,m}_k$ is the level curve
$Im{f^{n,m}_k}$. By the construction of the line complex, there is a closed union of edges $L$ of the Critical Graph $C$ cutting the critical graph
in two two halves which are, topologically, one the mirror image of the other. Hence $C$ is equivalent to its mirror image.

Let us consider the function $g^{n,m}_k=\overline{f^{n,m}_k(\overline{z})}$. They have the same critical values and the critical graph
$D$ of $g^{n,m}_k$ is the mirror image under complex conjugation $C$.
Hence $D$ and $C$ are equivalent and so -by the Riemann Existence Theorem- $f$ and $g$ are equivalent.
Since three critical points $z_0,z_2,z_{-2}$ are fixed by complex conjugation then $g^{n,m}_k=f^{n,m}_{k}$.
Moreover, the real axis is the line $L$ cutting $C$ into two equivalent halves.

We now prove $(iii)$.  To show that $z_{-1}> 1$, we look at the restriction of the Critical Graph on the line $L$, i.e. the real axis.
Since in the Critical Graph there is no edge connecting $z_{-1}$ and $z_2$, or connecting
$z_1$ and $z_{-2}$ then $z_{-1}>z_{-2}=1$ and $z_1<z_2=0$.
\end{proof}

\end{lem}
\begin{rem*}
Due to above Lemma \ref{lem:ktritronquee}(iii), all the critical points of the rational function $f^{n,m}_k$ are real. This property characterizes
a class of rational function which was much studied in relation of a conjecture by B. and M. Shapiro \cite{eremenko02}. Hence, $f^{n,m}_k$ belongs
to two important and distinct classes of functions, it is Belyi and its critical points are all real.
\end{rem*}

\subsection{$f_n^{n,n}$}
Here we tackle the explicit construction of the functions $f_n^{n,n}$. All these functions
are defined on $\bb{Q}$.
\begin{Def}
Let $A_n(z)$ be the rational functions with critical data
\begin{equation}
\lbrace (0,2n,0), (1,2n,1) , (\infty,2n,0) \rbrace \, , \; n \in \bb{N} \; . 
\end{equation}
We call it the n-Airy function. Its line complex is shown in Figure \ref{figure:an} below.
\end{Def}
Functions $A_n$ coincide - up to a fractional linear transformation- to the functions $f_h$ (case $h=n, q=3$)
in \cite{nevanlinna30} \S 2. They are the approximants of the solution of the Airy equation
$$
\lbrace f,z \rbrace =- 2 z \; .
$$
The coefficients of $A_n$ are given explicitely in \cite{nevanlinna30}. 
\begin{figure}[htbp]
\input{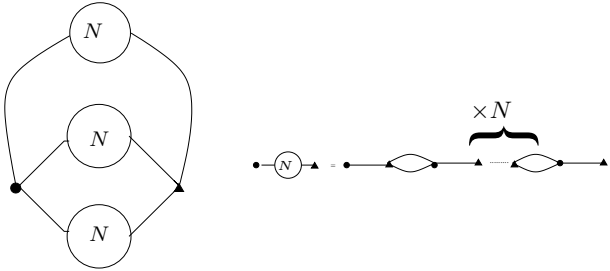}
 \caption{The line complex of the function $A_n$}
\label{figure:an}
\end{figure}
\begin{thm}\label{thm:fnnn}
$f_n^{n,n}(z)=A_n(f_0^{0,0}(z))$, $f_0^{0,0}=-2z^3+3z^2$. The critical data of $f_n^{n,n}$ are
\begin{equation}\label{eq:fnnncrit}
 \lbrace (0,4n+1,0) (1,4n+1,1), (\infty, 6n+2,\infty), (-\frac{1}{2},2n,1) , (\frac{3}{2},2n,0)  \rbrace \; .
\end{equation}
\begin{proof}
The function $A_n(f_0^{0,0}(z))$ has critical data (\ref{eq:fnnncrit}) so it satisfies
the same Hurwitz problem as $f_n^{n,n}$ (see Lemma \ref{lem:ktritronquee}).
To prove the Theorem it suffices to show that the two functions have
the same monodromy representation. This can be easily verified by composing the line complex of $A_n$ and of $f_0^{0,0}$.
\end{proof}
\end{thm}
The possibility of the non-trivial decomposition $f_n^{n,n}(z)=A_n(f_0^{0,0}(z))$ is due to the fact that the monodromy group
of $f_n^{n,n}$ is \textit{imprimitive} (Ritt's Theorem, see \cite{lando2004}).
Whether other functions $f_k^{n,m}$ have imprimitive monodromy group is an interesting combinatorial problem
that we do not address here.

Below are the Critical Graph of $f_n^{n,n}, n=0,\dots,3$. Here $z_0=w_0=1,z_{\pm2}=w_{\pm2}=e^{i\frac{2 \pi}{3}}$
and the Graph is the level curve $|f|=1$ \footnote{In the last picture, $1$ is a critical point of order $20$, hence close to $1$ the function is a constant
up to machine precision. This is the reason why the plot is not well-resolved.}.

\begin{minipage}[c]{.40\textwidth}
\begin{center}
\includegraphics[width=4cm]{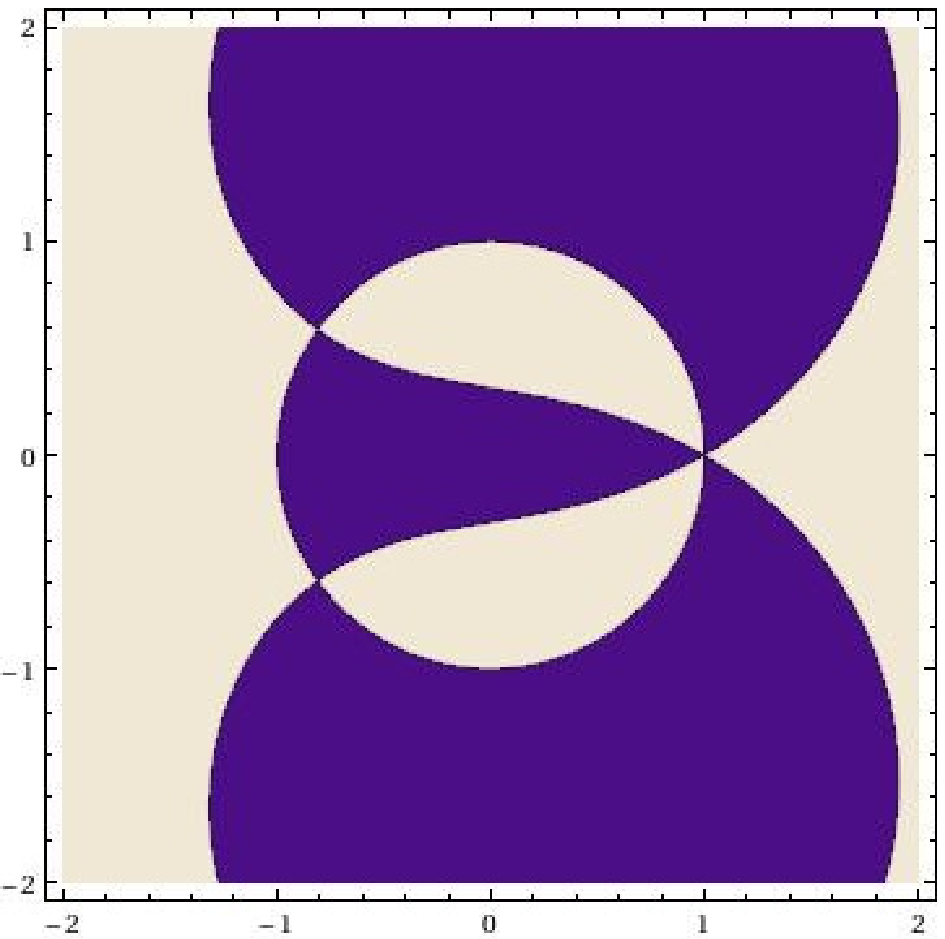}
\end{center}
\end{minipage} \hspace{1cm}
\begin{minipage}[c]{.40\textwidth}
\begin{center}
\includegraphics[width=4cm]{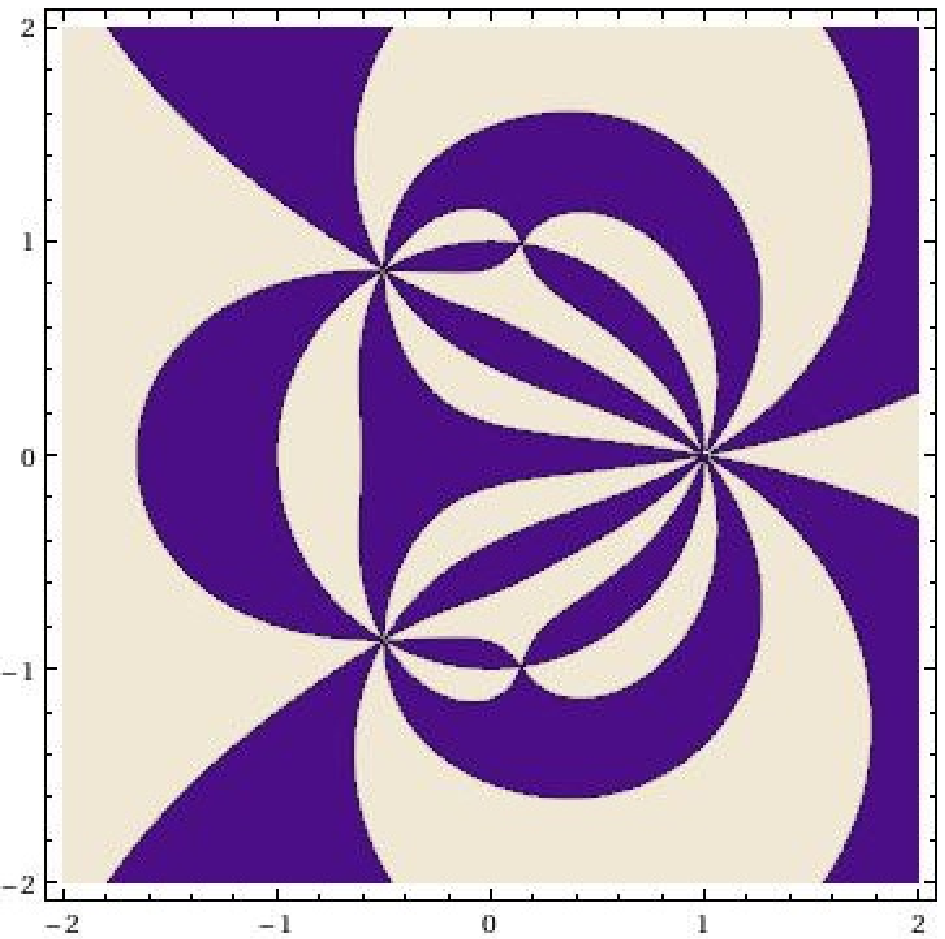}
\end{center}
\end{minipage}

\begin{minipage}[c]{.40\textwidth}
\begin{center}
\includegraphics[width=4cm]{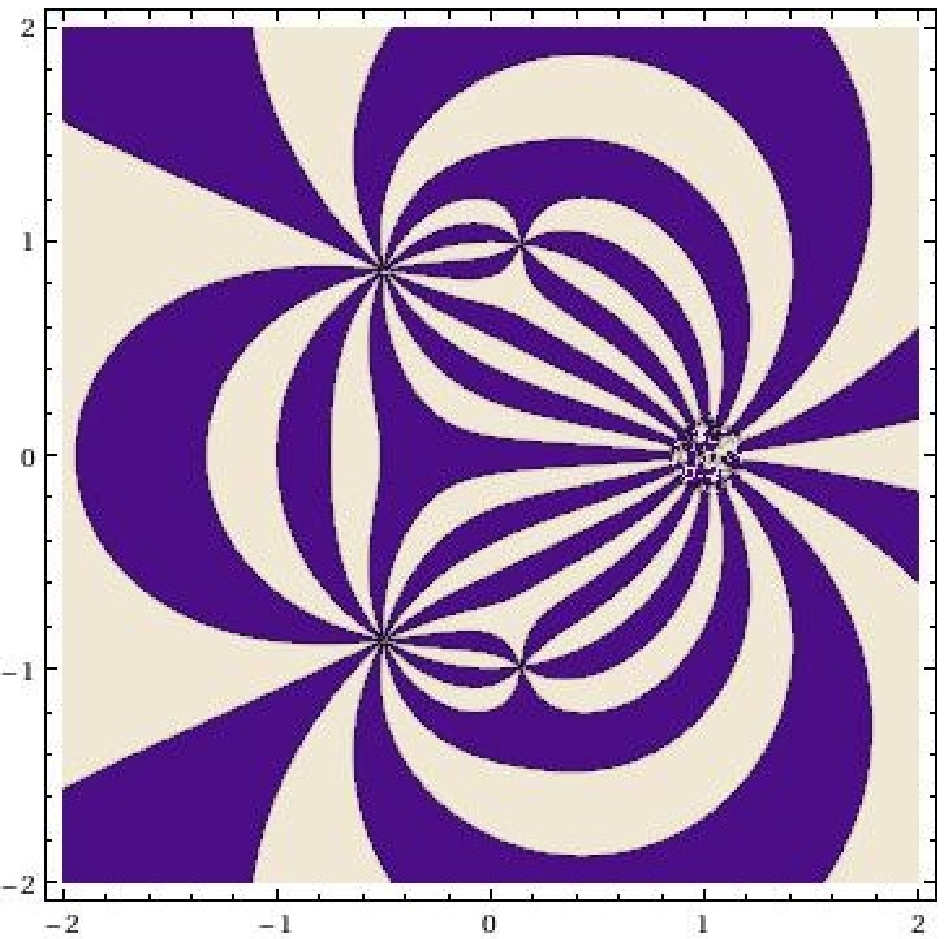}
\end{center}
\end{minipage} \hspace{1cm}
\begin{minipage}[c]{.40\textwidth}
\begin{center}
\includegraphics[width=4cm]{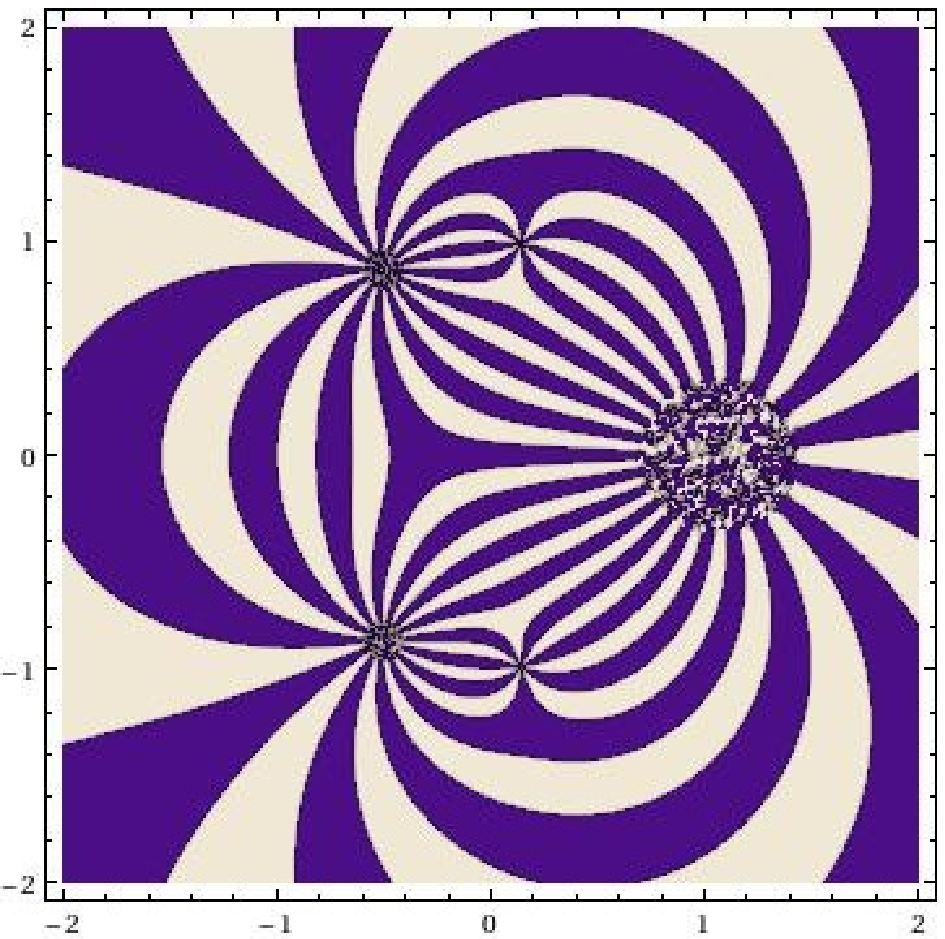}
\end{center}
\end{minipage}
\subsection{Numerical Experiments}\label{subsec:numerics}
We have computed numerically the functions $f^{n,m}_k$ for small $n,m,k$, by solving the corresponding Riemann problem.
To this aim -fixed $b_0=z_0=\infty,b_2=z_2=0,b_{-2}=z_{-2}=1$ - we solved the associated Hurwitz problem by looking
for admissible solutions of the system (\ref{eq:wedge}). For small $n,m,k$ conditions
(\ref{ineq:zz}) are enough to select a unique solution of the Hurwitz problem, which is the solution
of our Riemann problem.

Once the functions are calculated we normalize them according to general prescription (see Theorem \ref{thm:nevconvergence})
to assure the convergence of the $k \to \infty$ limit:
The sequence $f_k^{n,m}$ would converge to the function sought function $f^{n,m}$ up to
an affine transformation of the domain.

Due to Theorem \ref{thm:polesvalues}, for extracting the value of the corresponding pole of the tritronquee solution,
the function $f$ has to be normalized in such a way that $a_3=4, a_2=0$ and that $w_0=1$.

In practice we have computed the zero $z^*$ of the second derivative of $\lbrace f_k^{n,m},z \rbrace$ closest to $z=0$ and applied a dilatation to get
$\lbrace f_k^{n,m},z \rbrace =-2(4 (z-z^*)^3 - 2a_k^{n,m} (z-z^*) -28 b_k^{n,m}$. As $k \to \infty$, $a_k^{n,m} \to a^{k,m}$, the pole
of the integrale tritronquee with line complex ${\cal{D}}^{n,m}_k$.
\paragraph{Numerical computation of $f_k^{0,0}$}
To show how the method works, here we present $f_k^{0,0}, k\leq5$ and we postpone a more systematic presentation to a
subsequent publication.

To compute them, we have solved the corresponding Hurwitz problem.
Due to (\ref{eq:z+z}), system (\ref{eq:wedge}) reduces
to a system of polynomial equations in one variable, the location of the critical point $z_{-1}$.

Below we list the resultant of that system, which is the polynomial equation satisfied by $z_{-1}$.  We call $P_k$ corresponding polynomial.
\begin{eqnarray*}
 P_0(z) &=&-3+2z \, , \;  P_1(z) =-7+4 z , \\
P_2(z) &=& 693-2310 z+2772 z^2-1416 z^3+256 z^4 , \\
P_3(z) &=& 1859 - 6006 z + 6864 z^2 - 3328 z^3 + 576 z^4 , \\
P_4(z) &=&  -23056709+187166226 z-667288284 z^2+1371124664 z^3 \\
& & -1788975552 z^4+1536508416 z^5-868265344 z^6 \\
& & +311072256 z^7-64052736 z^8+5767168 z^9 , \\
P_5(z) &=& -16232365+129858920 z-454506220 z^2+913469840 z^3 \\
& &-1161895600 z^4+970042112 z^5-531587776 z^6 \\
& &+184365056 z^7-36708352 z^8+3194880 z^9 \; .
\end{eqnarray*}
All $P_k$ listed above are irreducible over the rational numbers.
For both $P_4$ and $P_5$ the Galois group is the full $S_9$.

Computed $z_{-1}$, the function $f^{0,0}_k$ is found by calculating the kernel of the critical evaluation map, whose terms are
polynomials in $z_{-1}$ with integer coefficients. Hence the number field defined by $P_k$ is also the field of definition of $f^{0,0}_k$.
% 
% From the degree of polynomials $P_k$ presented here and the many other we have computed, we propose the following Conjecture about the degree
% of the number field over which the functions $f^{n,n}_k$ are defined.
% \begin{conj}\label{conj:degrees}
% ... 
% \end{conj}
The approximations $a_k^{0,0}, 3\leq k \leq 6 $ of the pole $a^{0,0}$ are
\begin{eqnarray*}
a_3^{0,0}=-2,57 , a_4^{0,0}=-2,53 , a_5^{0,0}=-2,50 \; .
\end{eqnarray*}
The reader will convince herself that $a^{0,0}$ is the first pole on the negative axis as computed in \cite{joshi}, namely
$a^{0,0}=-2,38\dots$ \footnote{The corresponding solution
of the Bohr-Sommerfeld-Boutroux system is $\alpha^{1,1}=- 2.32$ \cite{piwkb}. This confirms our expectations (see Subsection \ref{subsec:wkb}).}.

We plot the Critical Graph of $f_k^{0,0}, k=0,\dots,5$. Here $w_0=1,w_{\pm2}=e^{i\frac{4 \pi}{5}}$
and the Graph is the level curve $|f_k^{0,0}(z)|=1$.

\begin{minipage}[c]{.40\textwidth}
\begin{center}
\includegraphics[width=4cm]{f000.eps}
\end{center}
\end{minipage} \hspace{1cm}
\begin{minipage}[c]{.40\textwidth}
\begin{center}
\includegraphics[width=4cm]{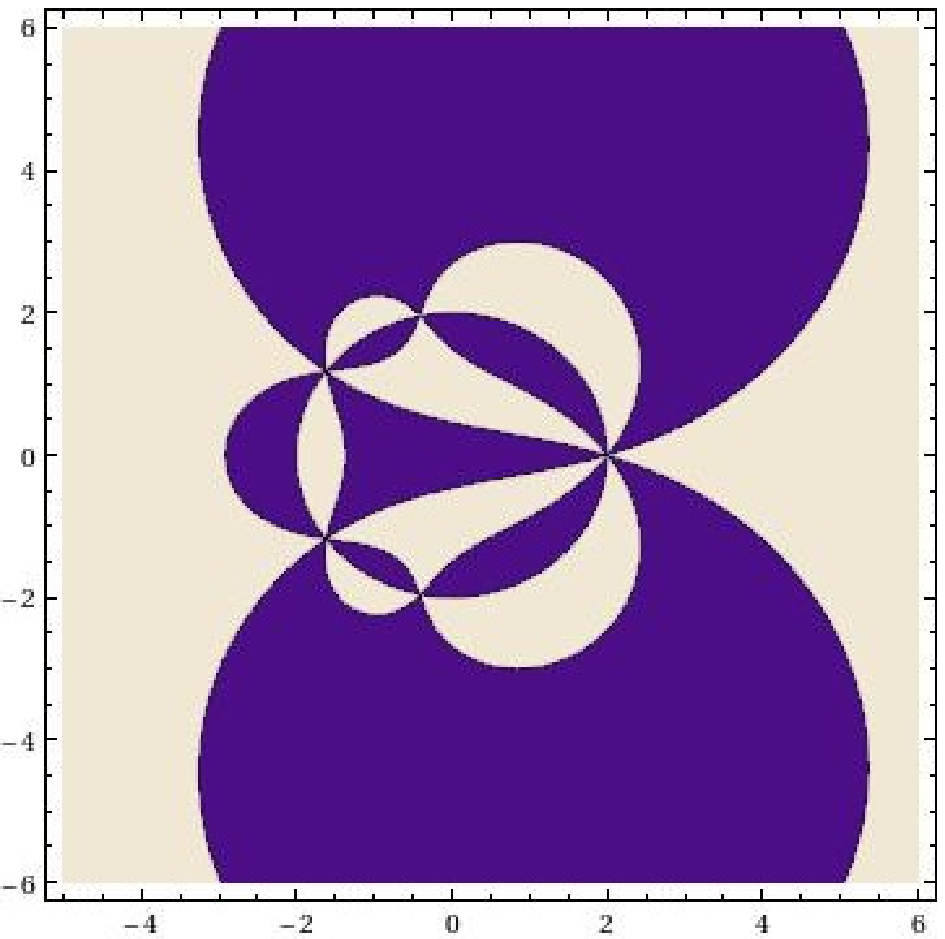}
\end{center}
\end{minipage}

\begin{minipage}[c]{.40\textwidth}
\begin{center}
\includegraphics[width=4cm]{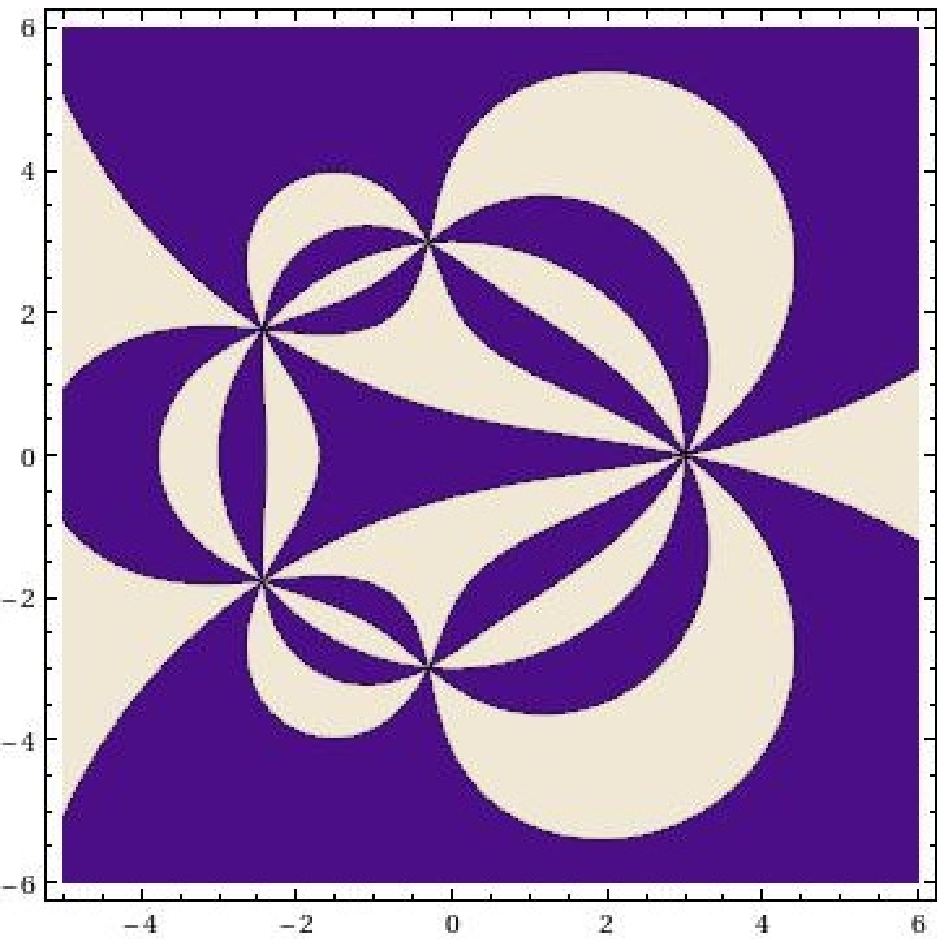}
\end{center}
\end{minipage} \hspace{1cm}
\begin{minipage}[c]{.40\textwidth}
\begin{center}
\includegraphics[width=4cm]{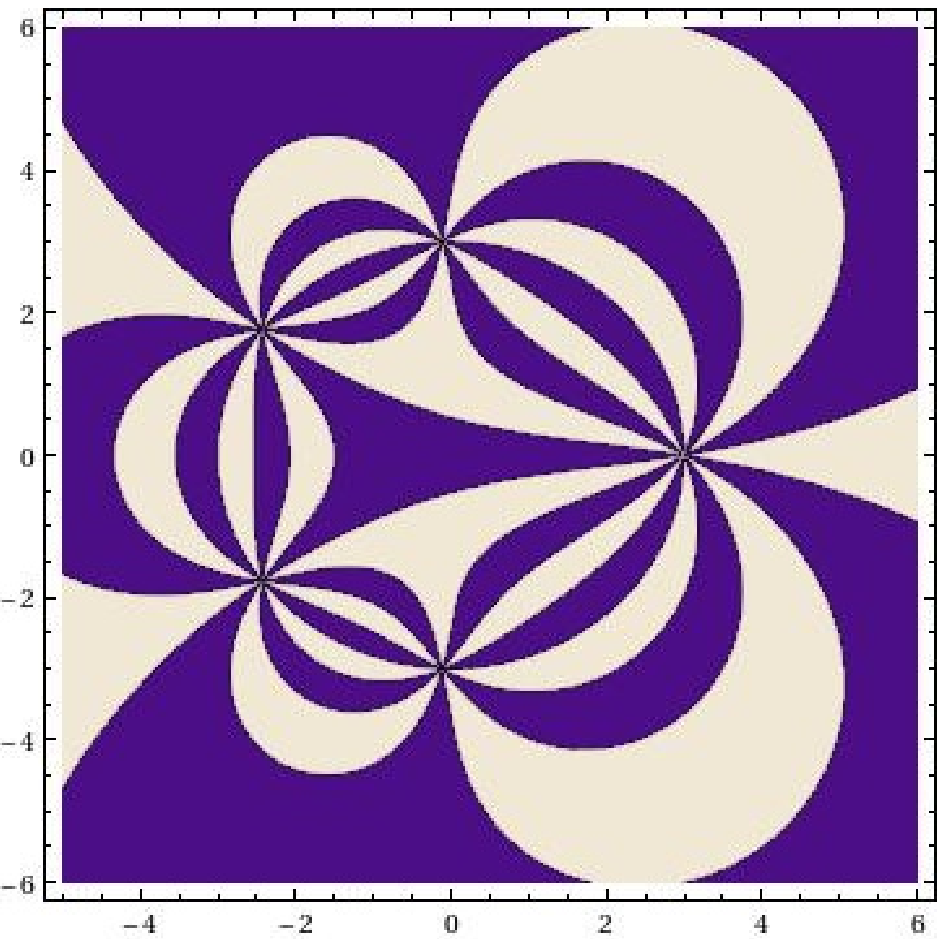}
\end{center}
\end{minipage}

\begin{minipage}[c]{.40\textwidth}
\begin{center}
\includegraphics[width=4cm]{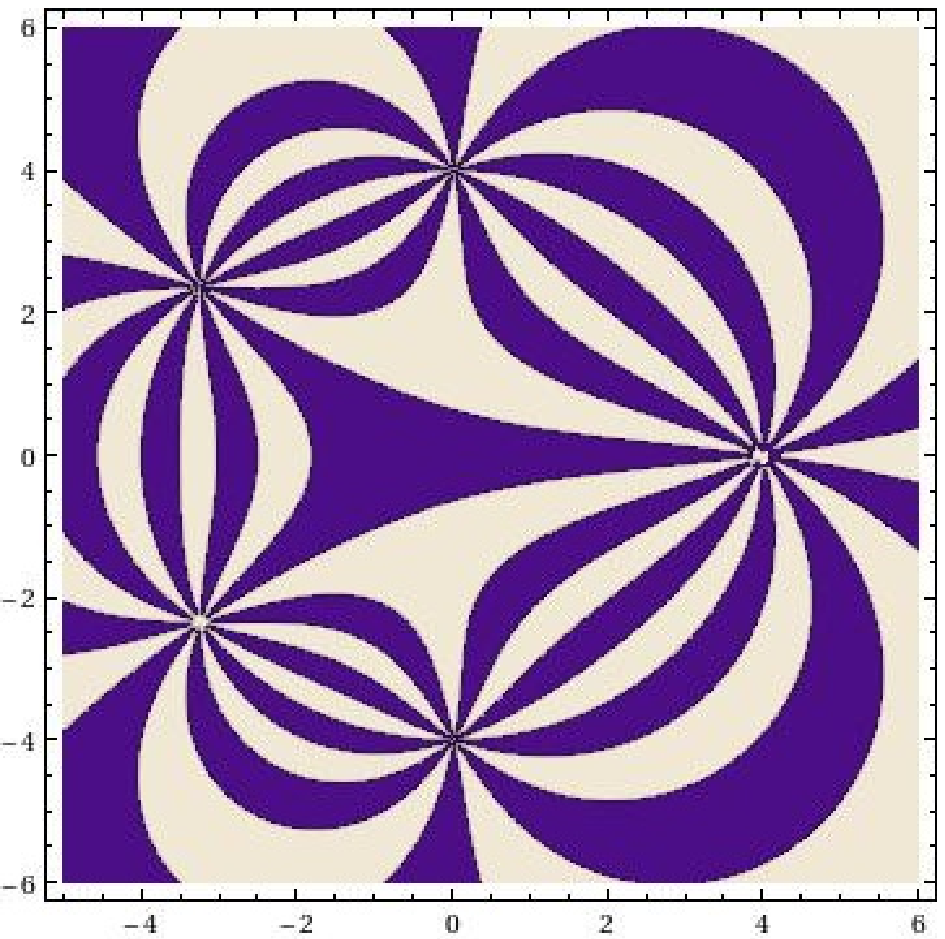}
\end{center}
\end{minipage} \hspace{1cm}
\begin{minipage}[c]{.40\textwidth}
\begin{center}
\includegraphics[width=4cm]{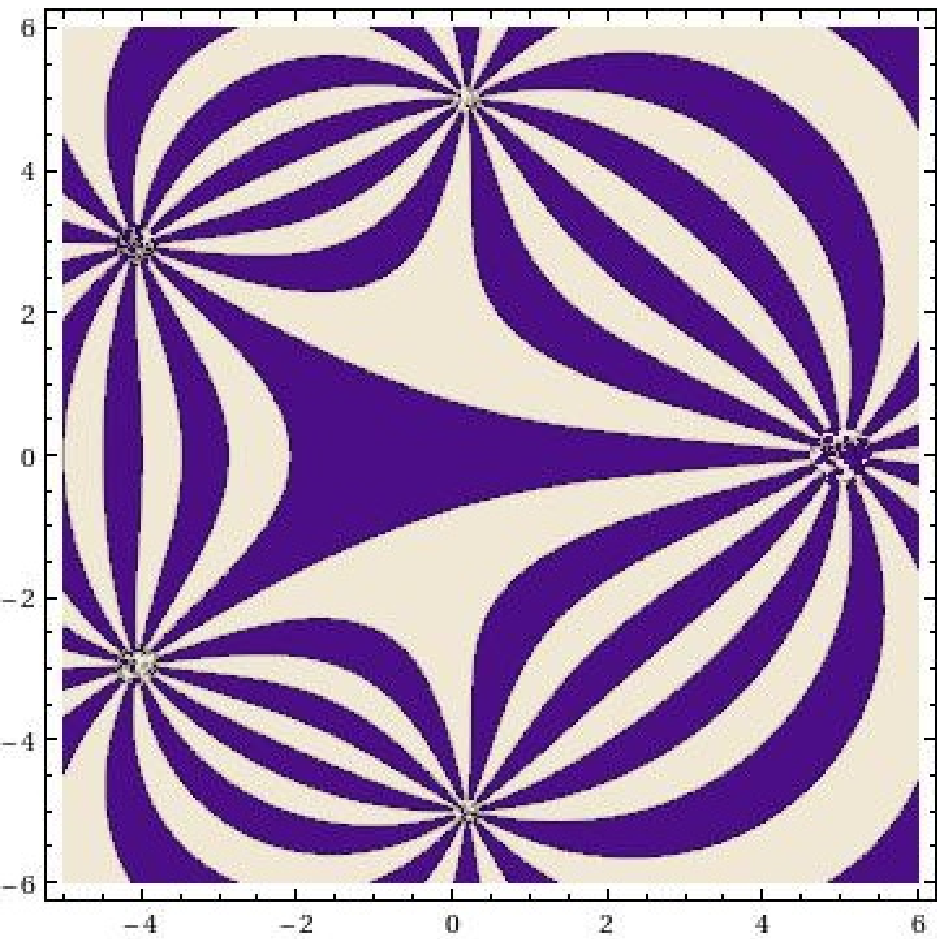}
\end{center}
\end{minipage}

\paragraph{$f_k^{1,0}$}

To give an example of poles that are not real, i.e. $n \neq m$, we plot $f_k^{1,0}, k=0,1$.

\begin{minipage}[c]{.40\textwidth}
\begin{center}
\includegraphics[width=4cm]{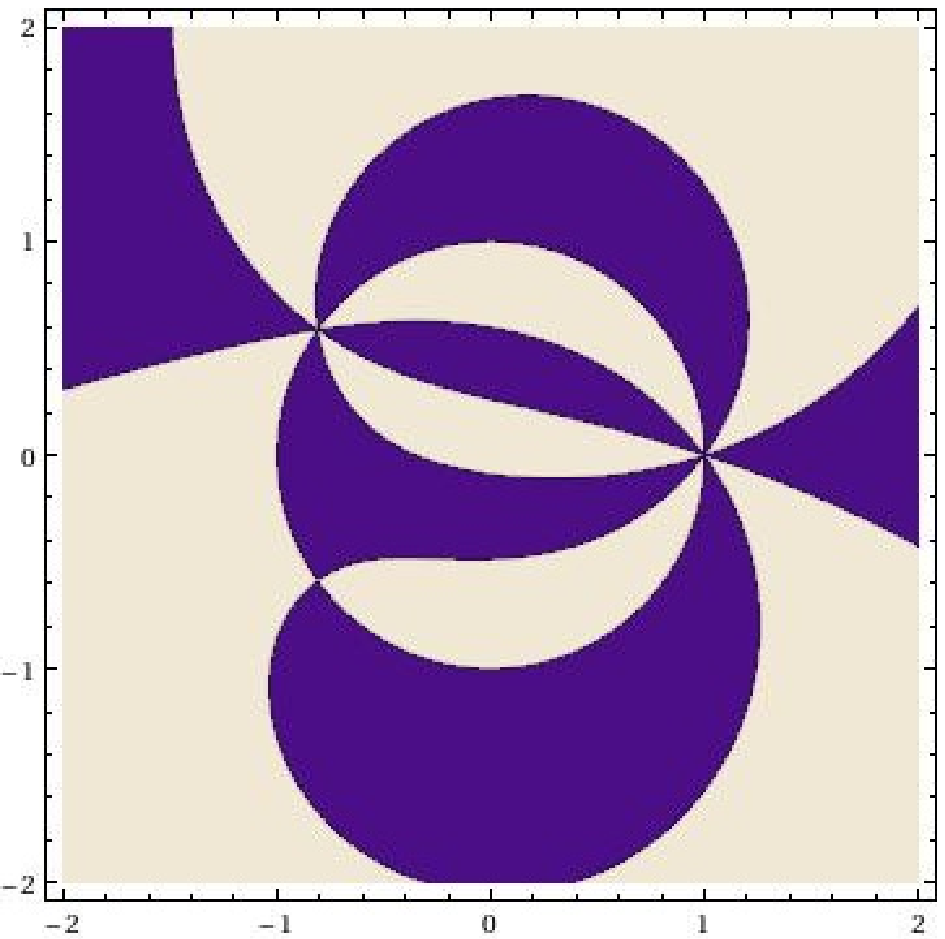}
\end{center}
\end{minipage} \hspace{1cm}
\begin{minipage}[c]{.40\textwidth}
\begin{center}
\includegraphics[width=4cm]{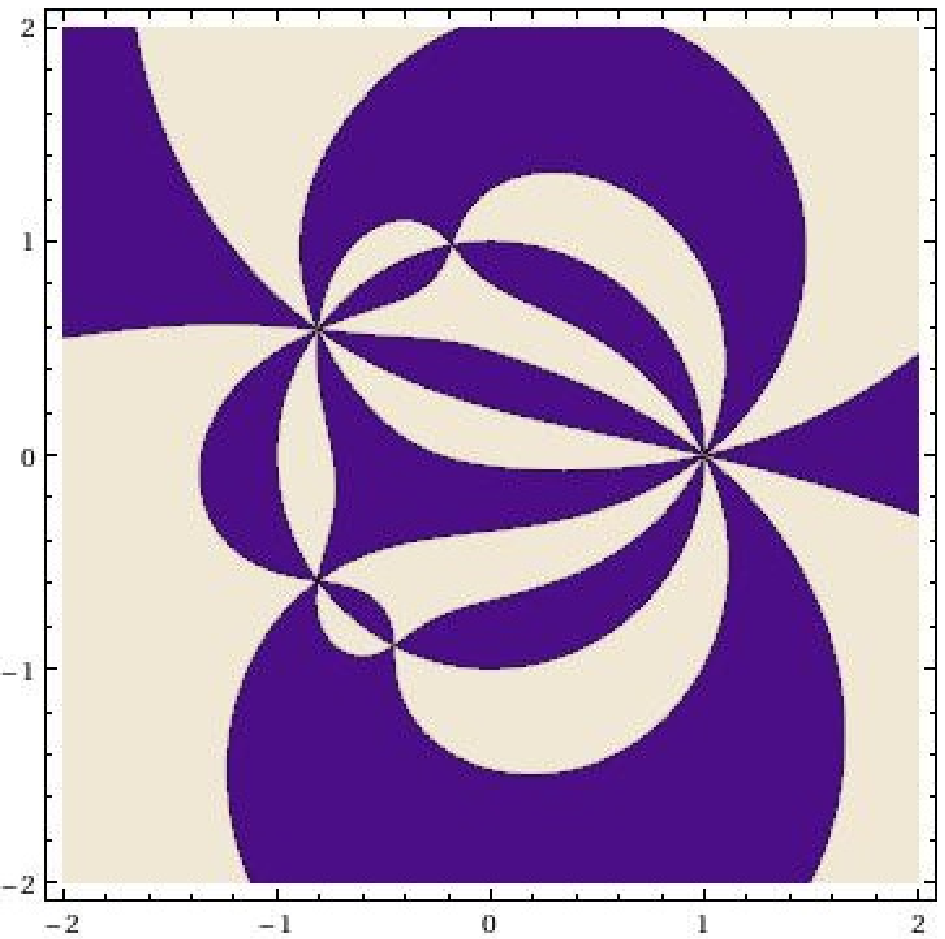}
\end{center}
\end{minipage}

\bibliographystyle{plain}
\bibliography{references}

\end{document}